\documentclass[%
superscriptaddress,
preprint,
 amsmath,amssymb,
 aps,
pre
]{revtex4-1}

\usepackage{graphicx}
\usepackage{dcolumn}
\usepackage{bm}
\usepackage{hyperref}

\allowdisplaybreaks 
\begin{document}


\title{No-boarding buses: Synchronisation for efficiency}
\author{Vee-Liem Saw}
\email{Vee-Liem@ntu.edu.sg}
\affiliation{Division of Physics and Applied Physics, School of Physical and Mathematical Sciences, 21 Nanyang Link, Nanyang Technological University, Singapore 637371}
\affiliation{Data Science and Artificial Intelligence Research Centre, Block N4 \#02a-32, Nanyang Avenue, Nanyang Technological University, Singapore 639798}
\author{Lock Yue Chew}
\email{lockyue@ntu.edu.sg}
\affiliation{Division of Physics and Applied Physics, School of Physical and Mathematical Sciences, 21 Nanyang Link, Nanyang Technological University, Singapore 637371}
\affiliation{Data Science and Artificial Intelligence Research Centre, Block N4 \#02a-32, Nanyang Avenue, Nanyang Technological University, Singapore 639798}
\affiliation{Complexity Institute, 61 Nanyang Drive, Nanyang Technological University, Singapore 637335}
%

\date{\today}

\begin{abstract}
We investigate a no-boarding policy in a system of $N$ buses serving $M$ bus stops in a loop, which is an entrainment mechanism to keep buses synchronised in a reasonably staggered configuration. Buses always allow alighting, but would disallow boarding if certain criteria are met. For an analytically tractable theory, buses move with the same natural speed (applicable to programmable self-driving buses), where the average waiting time experienced by passengers waiting at the bus stop for a bus to arrive can be calculated. The analytical results show that a no-boarding policy can dramatically reduce the average waiting time, as compared to the usual situation without the no-boarding policy. Subsequently, we carry out simulations to verify these theoretical analyses, also extending the simulations to typical human-driven buses with different natural speeds based on real data. Finally, a simple general adaptive algorithm is implemented to dynamically determine when to implement no-boarding in a simulation for a real university shuttle bus service.
\end{abstract}

\maketitle


\section{Introduction}

In many bus systems, bus bunching is a natural repercussion where an initially staggered configuration of buses ends up with multiple buses getting closer to each other \cite{Vee2019,Newell64,Chapman78, Powell83,Gers09,Bell10}. An intuitive mechanism that explains this phenomenon is that if there is some perturbation in the spacing between buses or the number of people at bus stops waiting to board a bus, then a bus may have to stop a bit longer to pick up people as well as allow people to alight. Consequently, the bus immediately behind catches up and by the time it reaches the bus stop, there may be less people for it to pick up. With less people, it need not stop as long as the bus ahead of it, allowing it to further catch up. Eventually, the leading bus has to pick up the majority of the people (which also increases the required stoppage time for it to allow more people to alight), whilst the trailing bus picks up relatively fewer people. These two buses inevitably end up bunching, due to the positive feedback which tends to slow down the leading bus and speed up the trailing bus. As these two bunched buses move in a single unit, people who just miss this pair would have to wait longer for the next bus (or for this same pair to return, if there are only two buses), compared to the situation where the buses are staggered such that the waiting time would only be a fraction of it. It is therefore undesirable for buses to bunch as it significantly delays passengers by inefficiently increasing their waiting time for a bus to arrive.

\subsection{Holding back buses, stop-skipping, deadheading}

There has been much effort carried out in attempting to address bus bunching. A common idea is to hold back some buses when bunching is perceived to be imminent, with extensions to adaptive and dynamic controls according to real-time situations \cite{Abk84,Ros98,Eber01,Hick01,Fu02,Bin06,Daganzo09,Cor10,Cats11,Gers11,Bart12,Moreira16,Wang18}, although holding back buses generally slows down the entire system. A holding strategy could be executed based on the goal of sticking to a schedule or maintaining the buses' headway.

In contrast to holding buses, some have considered the option of skipping bus stops to speed up a slow bus \cite{Li91,Eber95,Fu03,Sun05,Cor10,Liu13}, as well as deadheading where an empty bus is dispatched directly to a target bus stop \cite{Furth85,Furth85b,Eber95,Eber98,Liu13}. Deadheading may be theoretically viewed as a special case of stop-skipping \cite{Liu13}. In the stop-skipping strategy, a simple formulation would be to designate a stretch of bus stops to be skipped and passengers who wish to go there would alight at a bus stop just before that stretch. These passengers are forced to wait for another bus, of course, losing priority for seats to those already on board \cite{Eber95,Sun05}. The solution method usually involves numerically optimising some objective function, using nonlinear integer programming for instance. An interesting extension to stop-skipping was investigated by Ref. \cite{Sun05}: A bus would allow passengers who wish to alight at a bus stop (where it would otherwise skip) to do so, and if so it would then \emph{also allow boarding} at that bus stop. The performance of this policy, however, is quite susceptible to the passenger distribution patterns especially if every bus stop generally has people who would like to visit. Incidentally, some of these studies aimed to optimise both performance improvement for passengers as well as cost reduction for the transit operators \cite{Fu03,Cats11,Liu13}.

Alternatively, a study which combined holding and stop-skipping strategies found this to be undesirable as such tight controls may induce poorer performance \cite{Lin95}. Perhaps a worthy way forward is to have buses with wide doors or ``no doors'' \cite{WideDoor,Steward14,Geneidy17}, so that multiple passengers may simultaneously board and alight as compared to queuing through a single narrow entry. However, this would risk other issues like fare evasion. Another possible approach involves city planning where the locations of bus stops along the bus routes are engineered to facilitate efficiency of the bus system \cite{Tirachini14}. Some other work considered optimising some objective function, involving various algorithms to enhance the bus system \cite{Zhao08,Ceder11,Tang18}, as well as being data driven \cite{Wang17b}.

\subsection{Boarding limits: A no-boarding policy}

Buses can be sped up by limiting the number of people who are allowed to board the bus \cite{Del09,Del12,Zhao16}, i.e. a \emph{no-boarding policy}. An initial investigation on trying to speed up a bus by limiting the number of people allowed to board was carried out by Ref.\ \cite{Del09}, where they combined this with the holding strategy. The performance of the bus system was compared with only the holding strategy and without any form of control. It turns out that the inclusion of boarding limits is beneficial, as observed from their simulations. The setup there comprises buses serving a loop of bus stops, where they start from one terminal and end at that same place after completing the loop. The optimal actions to be taken by a bus when it is at a bus stop are determined by an objective function (solved numerically, subject to a number of constraints) that comprises waiting time of commuters at bus stops, extra time spent on buses due to holding, as well as additional time spent at bus stops when passengers are denied boarding or if capacity limit is reached. All buses are assumed to have identical speed, no overtaking is allowed, and they introduced some kind of a ``dummy bus'' as a boundary condition that ensures that all passengers are picked up, after all other buses have returned to the terminal.

This model which dealt with boarding limits was subsequently improved and expanded in Ref.\ \cite{Del12}, with a similar objective function but with many more constraints to satisfy. Once again, the objective function that determines the action of a bus when it arrives at a bus stop is solved numerically, this time without the inclusion of the ``dummy bus''. With a more elaborate model, it allowed for different finite bus capacities and different bus speeds. As before in Ref.\ \cite{Del09}, they found that the inclusion of boarding limits would help improve the system. Furthermore, the boarding limit strategy led to better comfort for passengers, since the passenger load would be spread out more evenly compared to having no such implementation. However, they noted that such improvements would depend on the conditions of the demand. In particular, the most significant benefit derived from the boarding limit strategy occurs when the headways between buses are short and when passengers demand is high.

Whilst the studies in Refs.\ \cite{Del09,Del12} are based on numerical optimisation of some objective function, Ref.\ \cite{Zhao16} attempted an analytical treatment on how boarding limits would help improve the bus system. However, their theoretical construction did not appear to consider alighting (i.e. commuters are continually packed into the buses). Unlike those in Refs.\ \cite{Del09,Del12}, they only studied the boarding limit strategy without involving the holding strategy.

\subsection{This paper: An analytical treatment on the no-boarding policy, with numerical simulations and validation based on a real university bus loop service}

Recently, work by Ref. \cite{Vee2019} modelled a system of buses serving a loop of bus stops as a system of coupled oscillators \cite{Syn03}, where each bus is represented by an autonomous oscillator. By doing so, bunching is derived as the ramification of phase synchronisation of these individual oscillators due to coupling at the bus stops, when demand exceeds a critical threshold. In fact, that framework revealed an insight on real buses where different human drivers have different natural speeds (frequency detuning): A system of buses having a greater degree of frequency detuning can help bunched buses to undo bunching since these differences in natural speeds allow a faster bus to break away from a slower one. As a result, such a system of buses does not exhibit sustained bunching during lull times when demand for service is low (below the critical threshold), but would contain clusters of sustained bunching or even be completely bunched during busy periods (above the critical threshold).

With the framework of bus bunching being described as a synchronisation phenomenon, Ref. \cite{Vee2019} pointed out a common property of such systems of coupled oscillators, viz. they can be periodically entrained by an external force to preserve a desired configuration, just like an ordinary clock can be periodically perturbed by a centralised high-quality time-keeping source to maintain its accuracy \cite{Syn03}. Here for this paper, we focus and expand on that property by studying this mechanism for preventing bus bunching: The implementation of a \emph{no-boarding policy} to achieve \emph{synchronisation of buses that maintains a reasonably staggered configuration}. According to this policy, a bus would always allow alighting but disallow boarding of new passengers at bus stops if certain criteria are met. The goal of the policy is to speed up a ``slow'' bus so that the bus immediately behind would not end up bunching with it. The rationale for no-boarding is justifiable to the passengers waiting at the bus stop, since the following bus is nearby and approaching soon.

This idea is in contrast to the holding back strategy which slows down the system instead and lengthens the time spent by passengers on the bus \cite{Abk84,Ros98,Hick01,Eber01,Fu02,Daganzo09,Cor10,Cats11}, since faster buses are held back in favour of slower ones in order to maintain their staggered headway. Moreover, if buses are held down at some point along the route, then designated waiting bays must have been planned and allocated beforehand otherwise they would obstruct traffic --- something unfeasible in dense city centres which are already short of sought after prime land. A no-boarding policy does not require any specific engineering, making it readily applicable in any existing city or landspace. Whilst this idea is similar to skipping bus stops \cite{Li91,Eber95,Fu03,Sun05,Cor10,Liu13}, it always allows passengers to alight at their desired destinations. Plus, the instruction to the passengers is arguably clearer: \emph{``Please do not board.''} This is direct, compared to the potentially confusing, \emph{``We are skipping this, $\cdots$, and that bus stops.''}.

In this paper, we derive a comprehensive analytical theory of the no-boarding policy which improves upon the work in Ref.\ \cite{Zhao16}, as we account for passengers alighting the bus both in our theory and simulations. The aim of the no-boarding policy here is to maintain a reasonably staggered headway between all buses serving a loop of bus stops. Our theory allows us to understand the mechanism of no-boarding leading to stable headway between buses, as well as deriving when such a no-boarding policy would \emph{fail or backfire}. As was noted in Ref.\ \cite{Del12}, this would work well for buses with short headway and \emph{high demand} from their simulations. In fact, we will find this to be true --- corroborating with Ref.\ \cite{Del12}, and are actually able to work out analytically that the no-boarding policy backfires in the lull period defined in Ref.\ \cite{Vee2019}, due to buses having different natural speeds. The full mathematical derivation for this critical transition where no-boarding fails is given in Ref.\ \cite{Chew2020}. This paper then concludes with extensive simulations based on parameters measured from a real university campus loop shuttle service, where buses serve a loop of bus stops continually with no start/stop terminal. Live data for those buses can be found on a website that is listed as Additional Information at the end of this paper. In contrast to Refs.\ \cite{Del09,Del12}, we allow overtaking in our theoretical analysis and simulations as this is commonly observed in the university shuttle bus loop service \cite{Vee2019}.

\subsection{Outline and organisation of this paper}

After some preliminaries in the next section, we proceed to investigate how such a no-boarding policy can significantly improve passenger service. To do so, we carry out \emph{exact analytical calculations} in the case where buses move with identical natural frequency (which is the way forward, when self-driving buses become the norm in the near future \cite{NTUautobuses}) to show that the average waiting time of passengers can be reduced dramatically compared to a system with no such interference policy where all buses end up bunching. This is because the no-boarding policy ensures that the buses are reasonably spaced out and not bunch. In fact, two possible no-boarding policies are explored, by looking at the phase difference of a bus from the bus immediately ahead (Section 3) or immediately behind (Section 4). Our analytical results provide clear evidence on the efficacy of this mechanism and how exactly improvement is achieved, even though it may appear a bit simplified under our assumptions.

To illustrate and test our analytical results, we carry out extensive simulations which turn out to agree very well with the analytical results. These simulations also provide further insights into the actual behaviour of the bus system under the no-boarding policy. Apart from that, we apply this no-boarding policy to a simulation of a bus system having $M=12$ staggered bus stops in a loop, served by $N$ buses moving with \emph{different natural frequencies}, in Section 5. This setup is modelled after a university shuttle bus service \cite{Vee2019} to see how the no-boarding policy would pan out in a real human-driven bus system, instead of programmable self-driving buses with identical natural frequency. In addition, we describe a simple adaptive algorithm which dynamically determines what angle (phase difference) to implement the no-boarding policy, in a bus system where the rates of people arriving at the various bus stops differ and are stochastic in Section 6. Those real parameters used in our simulations are measured from the university shuttle bus service \cite{Vee2019}, thereby providing support that our analytical theory's results are indeed applicable to realistic systems with varying demand for service. The simplicity of this adaptive algorithm makes it general enough to be directly implemented in generic bus systems, which can be further improved and fine-tuned towards specific systems especially given prior knowledge and empirical data of the real systems.

\begin{figure}
\centering
\includegraphics[width=16cm]{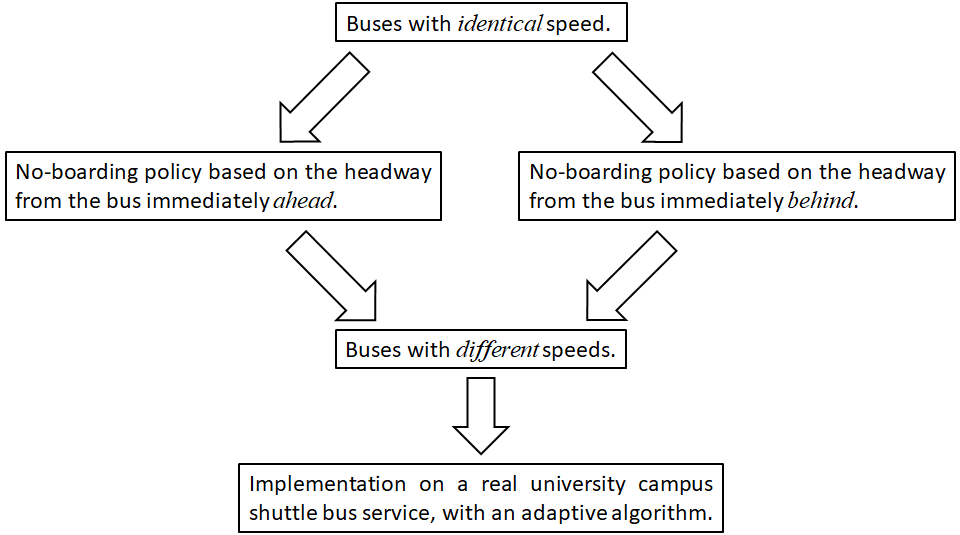}
\caption{Outline and organisation of this paper.}
\label{fig1}
\end{figure}

A flow diagram in Fig.\ \ref{fig1} summarises the layout and organisation of this paper.

\section{Preliminary}

Consider a bus loop system comprising $N$ buses serving $M$ staggered bus stops around the loop. This loop can be mapped isometrically (preserving distances) to the unit circle. Each bus $i$, where $i\in\{1,\cdots,N\}$, moves with its natural (angular) frequency $\omega_i=2\pi f_i=2\pi/T_i$. With bus stop $j$ present in the loop, where $j\in\{1,\cdots,M\}$, each bus $i$ must spend a stoppage $\tau_{ij}$ to let passengers board or alight. We consider the process of boarding new passengers to occur after all passengers on the bus who wish to alight have done so. This is similar to the elevator scenario where new users would only enter the elevator \emph{after} everyone currently inside (who intends to exit at that level) has completely vacated. Hence, this stoppage $\tau_{ij}$ is due to $P_j:=$ number of people at bus stop $j$, $Q_{ij}:=$ number of people on bus $i$ who are alighting at bus stop $j$, and $l:=$ loading rate of passengers onto the bus which we assume is the same as the unloading rate of passengers off the bus, i.e. $\tau_{ij}=(P_j+Q_{ij})/l$. On average, especially if the $M$ bus stops are equally popular, we have $P_j\sim Q_{ij}$ (i.e. the number of people originating from a bus stop is comparable to the number of people heading to that bus stop) so that $\tau_{ij}\sim2P_j/l$. However, this is not necessarily true especially during peak hours due to a bias where people from one region (say homes in the morning) tend to head towards another part of the bus route (offices, business district). Incidentally, $P_j$ depends on the time headway $\Delta t_{ij}$ between bus $i$ and the bus immediately ahead (temporal phase difference) together with the average rate of people arriving at bus stop $j$ to wait for a bus, denoted by $s_j$, i.e. $P_j=s_j\Delta t_{ij}$. This implies that the larger the headway between bus $i$ and the bus immediately ahead, the more people it has to pick up and consequently the longer it has to stay at bus stop $j$ --- slowing it down even more and further increasing $\Delta t_{ij}$. This positive feedback inevitably leads to the perennial and notorious problem of bus bunching which significantly reduces the bus system's efficiency and lengthens the waiting time for passengers.

For simplicity of the subsequent analysis to present the key ideas, we first focus on the system of $N=2$ buses serving $M=1$ bus stop in a loop and the two buses have the same natural (angular) frequency $\omega=2\pi f=2\pi/T$ (this $T$ excludes stoppage at the bus stop). In \emph{steady state}, we can approximate $P_j=Q_{ij}$ giving $\tau_{ij}=2P_j/l$, because there is only one bus stop and every passenger who boards there must alight there after one loop. We should emphasise that $P_j$ and $Q_{ij}$ generally fluctuate instead of being equal, but are on average equal over the time scale of several $T$. In particular, $P_j=s_j\Delta t_{ij}$ is true but $Q_{ij}$ depends on $\Delta t_{ij}$ from the \emph{previous} time when these people boarded instead of the present $\Delta t_{ij}$. This would lead to a recursive relation between $\Delta t_{ij}$ for this round in terms of $\Delta t_{ij}$ for the previous round. Nevertheless over several $T$ so that $P_j\sim Q_{ij}$, we have
\begin{align}
\tau_{ij}=2k_j\Delta t_{ij},
\end{align}
with $k_j:=s_j/l<1$ being parameters that determine the strength of coupling between the buses. This equation would hold in general for any $N$ and $M$, if we have the condition $P_j=Q_{ij}$ to be true. We adopt this assumption in our subsequent theoretical analysis for general $N$ and $M$ to enable a tractable analytical treatment towards gleaning useful insights on the bus system. In that case, we may consider each bus stop as equally popular and so all $s_j=s$, with $k:=s/l$. In our simulations, we find that the steady state results turn out to match with the analytical results thereby justifying the simplification and approximation made here. Later in Section \ref{realworld}, we drop such simplification that all bus stops have the same $s_j=s$ in our simulations and find that the key results from the analytical theory carry over to realistic systems based on real-world parameters.

The setup here is motivated and modelled from our Nanyang Technological University's (NTU) campus loop shuttle bus service. A bus typically completes a loop within 20 minutes, serving $M=12$ reasonably staggered bus stops. (More details in Section \ref{NTUsimulations}.) Bus bunching is a common occurrence for this loop service \cite{Vee2019,NTUnews}. There is no traffic light on campus, hence work in Ref.\ \cite{Vee2019} found that the primary mechanism for bus bunching in such a system is due to the coupling amongst buses from stopping at bus stops to embark and disembark passengers. There is no start/end terminal, with the buses continuously looping around the route, though buses occasionally pull out and new buses add to the service. The interarrival times between two buses are quite frequent, viz. as short as a couple of minutes during peak hours to about $10$ minutes during off-peak hours. Hence, if a bus implements no-boarding, it would cost only several minutes before the next bus arrives.

We assume that there are enough buses going around to meet the demand, such that the finite bus capacity of the buses is not reached. In other words, a passenger is not able to board a bus only because the bus implements no-boarding. Also, it may occur that a passenger is denied boarding by more than one bus consecutively, and the extended waiting time is tracked by the simulations. It turns out that in steady state, the overall average waiting time would be better than normal buses and follows the analytical results derived here.

In future work, one may move on with imposing finite bus capacity, traffic conditions, different loading/unloading rates, and other real-world features of specific bus systems. The effect of traffic conditions would then extend the analysis of how the no-boarding policy would fare in situations where buses may bunch due to uncertainty in arrival times and fluctuations due to traffic congestion. The aim of this paper is to elucidate what a no-boarding policy can do, at least in an idealised situation.

In comparison with other strategies like holding, stop-skipping and deadheading, Refs.\ \cite{Del09,Del12} have shown via their simulations that the inclusion of the no-boarding policy helps improve the performance of the bus system, as compared to only implementing the holding strategy. We further explore the performance of the no-boarding strategy and the holding strategy elsewhere in Ref.\ \cite{Vee2019d} using a reinforcement learning approach. In our setup here where every bus stop has people boarding as well as alighting, the idea of stop-skipping would necessarily force a fair amount of passengers to alight at an earlier stop and then board another bus just to complete their journey. This certainly increases their total waiting and travel times, with the inconvenience of getting off and then on a new bus. Besides that, the idea of deadheading implies the dedication of a bus for this purpose which reduces the number of buses serving other bus stops. Since every bus stop has people who need service, there is no benefit from deadheading here. Instead, deadheading would be useful if there is some particular bus stop that serves as a hub, i.e. with significantly higher demand for service compared to other bus stops. This is consistent with other work showing that it is not quite desirable as it has its own problems (see second paragraph of Section 1A).

\section{No-boarding policy: Looking ahead}

\subsection{Analytical theory for \texorpdfstring{$N=2$}{N=2} buses serving \texorpdfstring{$M=1$}{M=1} bus stop}

\subsubsection{Stoppage duration, \texorpdfstring{$\tau$}{stoppage}}

In the bus loop system comprising $N=2$ buses serving $M=1$ bus stop, suppose the two buses never bunch due to some implemented policy, i.e. their spatial headway or \emph{phase difference} $\Delta\theta$ on the unit circle is never equal to $0$ or $2\pi$ radians, but is maintained to remain within a small range of ideal angles. In such a steady state, both buses must, on average, spend an \emph{equal} stoppage $\tau$ at the bus stop, otherwise one bus systematically stopping shorter than the other would lead to bunching. Therefore, one revolution by each bus takes a time of $T+\tau$, with a total of $s(T+\tau)$ people arriving at the bus stop. Each of the two buses picks up half this number of people, over an average time interval of $\tau/2$ (since the other $\tau/2$ is for people to alight) at a rate of $l$. This gives the relationship:
\begin{align}
\frac{1}{2}s(T+\tau)&=\frac{1}{2}l\tau\\
\tau(k, T)&=\frac{kT}{1-k}\label{taufull}\\
\overline{\tau}(k)&=\sum_{n=1}^{\infty}{k^n}.\label{tau}
\end{align}
The geometric series expansion is valid since $k<1$, and looks like a convenient way of expressing $\overline{\tau}(k)$ [also for the general case with $N$ buses, cf. Eq.\ (\ref{tauN})]. Here, we have defined $\overline{\tau}$ equals to $\tau$ per unit $T$ as a normalisation unit. In other words, when the two buses are in such a steady state where they both spend an equal average stoppage $\tau$ at the bus stop, then $\overline{\tau}$ is given by Eq.\ (\ref{tau}), which depends only on $k:=s/l$. The number of people picked up by each bus is then
\begin{align}
L(s, l, T)=\frac{1}{2}l\tau.\label{people}
\end{align}

Without any intervention, the two buses cannot remain staggered because such a configuration is unstable. This would be exemplified by systems with the buses moving at different natural frequencies, where frequency detuning leads to periodic bunching during lull periods, otherwise sustained bunching is the outcome during busy periods \cite{Vee2019}. To prevent bunching, some form of intervention has to be executed. A way to go would be to implement a \emph{no-boarding policy} when the phase difference $\Delta\theta$ deviates too much from the staggered configuration. To be precise, if bus $i$ is at bus stop $j$, then bus $i$ checks its phase difference with respect to the bus immediately ahead, $\Delta\theta$. (Alternatively, it can check its phase difference with respect to the bus immediately behind, discussed in the next section.) Since bus $i$ is stationary at the bus stop, then $\Delta\theta$ is non-decreasing (it can be momentarily unchanging if the bus immediately ahead happens to be at a bus stop, for a system with $M>1$ bus stops). If $\Delta\theta$ exceeds some critical angle $\theta_0$, then according to this \emph{no-boarding policy}, bus $i$ disallows any further boarding to halt its delay with respect to the bus immediately ahead. Of course, if there are still people on bus $i$ who need to alight, then bus $i$ would wait until they have completely alighted.

\subsubsection{Lower bound to \texorpdfstring{$\theta_0$}{angle}: \texorpdfstring{$\theta_\textrm{min}$}{lower bound}}

\begin{figure}
\centering
\includegraphics[width=16cm]{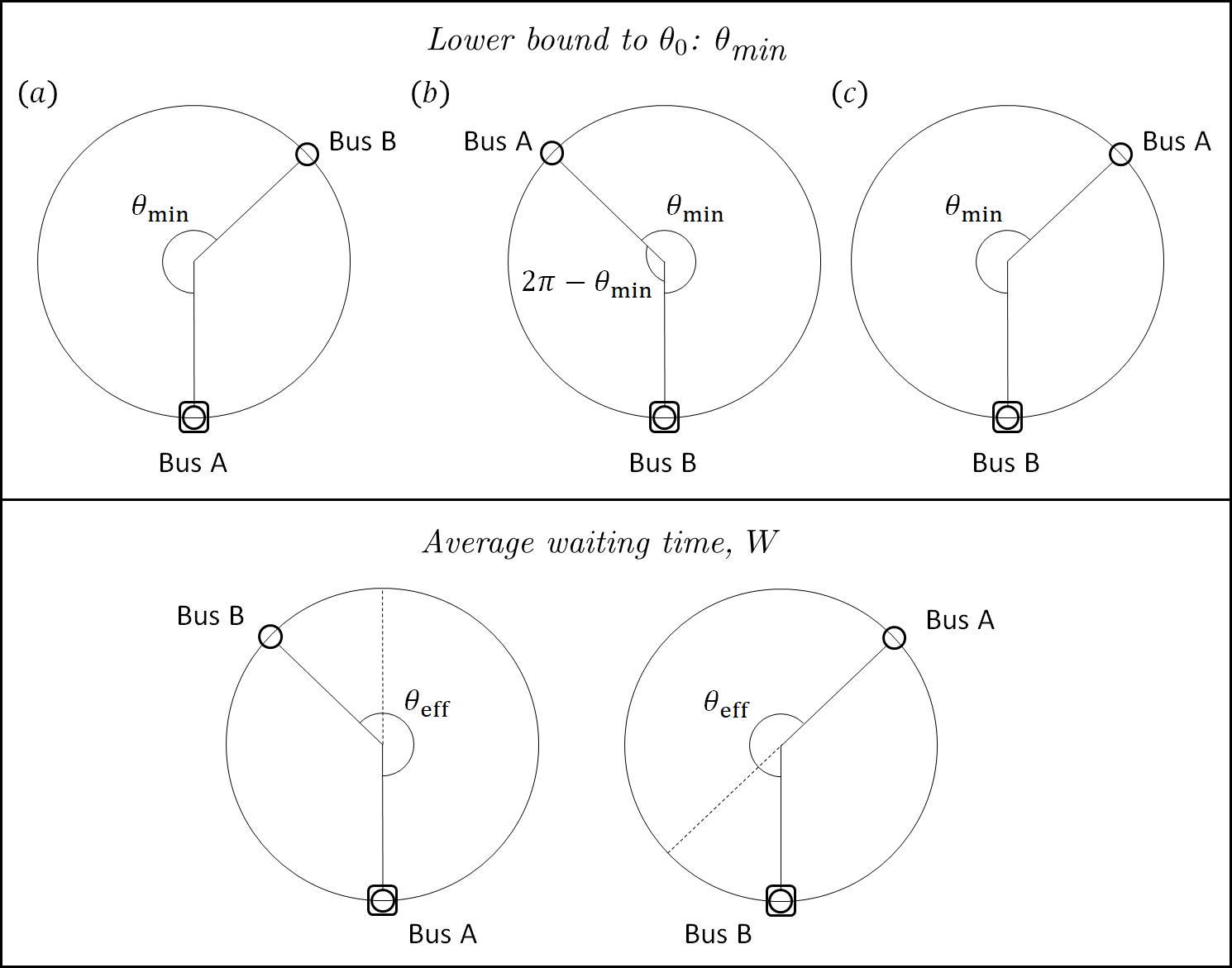}
\caption{$N=2$ buses serving $M=1$ bus stop in a loop. The buses travel clockwise, and implement the no-boarding policy if $\Delta\theta>\theta_0$. Top panel: Shown here is the situation where $\theta_0$ is the lower bound $\theta_\textrm{min}$, before the system is picking up people slower than people arriving at the bus stops. Bottom panel: In steady state, the phase difference $\Delta\theta(t)$ fluctuates around the effective angle $\theta_\textrm{eff}$, which is bounded by $\theta_0$ (i.e. $\theta_\textrm{eff}<\theta_0$). On average, bus A and bus B would stop over the same duration $\tau$ and pick up the same number of people $L$. Since bus B is lagging due to its large phase difference with respect to bus A, a portion of $(1-\pi/\theta_\textrm{eff})\times100\%$ of the people waiting for bus B would have to wait longer to board bus A instead, because bus B would have implemented the no-boarding policy when $\Delta\theta>\theta_0$. Hence if $\theta_\textrm{eff}$ is closer to $\pi$, then fewer people would have to be denied boarding by bus B. A closer-to-antipodal configuration would improve the overall average waiting time, given by Eq.\ (\ref{waiting}).}
\label{fig2}
\end{figure}

Note that there is a lower bound to $\theta_0$ where no-boarding is implemented if $\Delta\theta>\theta_0$. This lower bound arises, because if $\theta_0$ is too small, then the buses would too frequently disallow boarding whenever $\Delta\theta>\theta_0$ with the repercussion that they are \emph{not picking up people faster than people arriving at the bus stop} --- leading to an unbounded growth of people at the bus stop waiting for service. We can calculate the lower bound to $\theta_0$, denoted as $\theta_\textrm{min}$ as follows. In Fig.\ \ref{fig2}(a) (top panel), bus A implements the no-boarding policy since it sees $\Delta\theta=\theta_\textrm{min}$, and leaves the bus stop. Fig.\ \ref{fig2}(b) (top panel) is the moment when bus B just arrives at the bus stop, and the phase difference remains as $\theta_\textrm{min}$ --- strictly speaking, since bus B is now at the bus stop, it is bus B that measures its phase difference with respect to bus A. Therefore, the phase difference is $\Delta\theta=2\pi-\theta_\textrm{min}<\pi$. As bus B remains at the bus stop, bus A advances until in Fig.\ \ref{fig2}(c) (top panel) where $\Delta\theta$ grows to $\theta_\textrm{min}$. If the time taken by bus A to go from (b) to (c) is less than $\tau$, then bus B has to implement no-boarding and stop for a duration \emph{less} than $\tau$ given in Eq.\ (\ref{taufull}). Consequently, bus B picks up \emph{less} than $L$ people given in Eq.\ (\ref{people}) --- monotonically creating a growing list of disgruntled people waiting for service. This $\theta_\textrm{min}$ is the critical lower bound such that bus B is able to spend exactly $\tau$ to be able to pick up the required $L$ number of people. Since the time interval from (b) to (c) is $(2\theta_\textrm{min}-2\pi)/\omega$ (the time taken by bus A to travel) and this equals to $\tau$ (the stoppage duration of bus B), we get
\begin{align}
\frac{2\theta_\textrm{min}-2\pi}{\omega}&=\frac{2\pi\overline{\tau}}{\omega}\\
x_\textrm{min}(k)&:=\frac{\theta_\textrm{min}}{2\pi}=\frac{1}{2}\left(1+\overline{\tau}(k)\right)\label{lowerbound},
\end{align}
where we have defined $x_\textrm{min}:=\theta_\textrm{min}/2\pi$ being a normalisation unit. We find that this minimum angle depends on $k$. If demand is higher, then the lower bound to the angle for implementing no-boarding is larger, i.e. the buses should be given greater leeway and duration to actually let people board since there are more people demanding service.

\subsubsection{Average waiting time, \texorpdfstring{$W$}{waiting}}

An important parameter which quantifies the efficiency of a bus system is how long a passenger takes to complete the journey, i.e the total trip time, which includes the waiting time at the bus stop and the time spent on the bus. Without the no-boarding intervention, bus bunching is inevitable, and so for the $N=2, M=1$ system, the bunched pair is effectively one large bus with twice the capacity of a single bus and twice the loading/unloading rate. If the system's two buses can ``magically'' remain antipodally staggered, the average waiting time would be halved. It is easy to see that if an unlucky person barely misses a bus: in the bunched case the waiting time is roughly one full revolution; whereas in the antipodally staggered case it is only about half a revolution.

Note that since the no-boarding policy always allows passengers who wish to alight to do so, it never extends their time spent on the bus. In other words, with the total trip time being the sum of the waiting time at the bus stop and the time spent on the bus, since the latter is unaffected, \emph{an improvement on the waiting time at the bus stop directly implies an improvement in the total trip time}. This is in contrast to the holding strategy where there is a trade-off between waiting time and time spent on the bus: If a bus is held down, then the time spent on the bus is lengthened for those passengers on the bus.

We can work out the average waiting time more precisely, where people are assumed to be arriving at the bus stop uniformly over time at a fixed rate of $s$. These people should obey the \emph{first-come-first-served} rule, i.e. the person who arrived at the bus stop earlier would be ahead on the waiting list to board the bus, after everybody on the bus has alighted. In addition, a bus is assumed to have only one door for one person to alight or board at a time, at a rate of $l$. This simplifies our calculations for the average waiting time of people at the bus stop for a bus to arrive, as it eliminates the possibility for multiple passengers simultaneously alighting or boarding. With one door, it is clear that the ordering on the waiting list of people is well-defined and a person ahead of the list necessarily gets up the bus first.

When the no-boarding policy is implemented at $\Delta\theta>\theta_0$, the (so-called ``slower'') bus at the bus stop disallows further boarding and leaves. Subsequently, the other (so-called ``faster'') bus would stop at the bus stop and the phase difference approaches the antipodal configuration. After this ``faster'' bus leaves, the cycle repeats with the ``slower'' bus stopping and the phase difference widening. The \emph{effective phase difference} between this pair of buses is $\theta_\textrm{eff}$, where the actual phase difference $\Delta\theta(t)$ fluctuates around $\theta_\textrm{eff}$ over the long term. The implemented angle for no-boarding $\theta_0$ would bound $\Delta\theta(t)$ and so $\theta_\textrm{eff}<\theta_0$. Occasionally though, $\Delta\theta(t)$ can exceed $\theta_0$. This happens when there are still people on the bus who wish to alight and the bus must oblige, even though $\Delta\theta(t)>\theta_0$. Through numerical simulations (next subsection), we find that $\theta_\textrm{eff}$ is reasonably represented by the \emph{median} of the buses' phase difference in steady state.

In the bottom panel of Fig.\ \ref{fig2}, bus B is the ``slower'' one due to its large phase difference with respect to bus A. As a result, bus B would implement the no-boarding policy when $\Delta\theta>\theta_0$, and people after that would have to wait further to board bus A. On the other hand, bus A always picks up everybody. To calculate the average waiting time for people at the bus stop before being able to board, we consider the waiting times to board bus A and bus B \emph{separately}. Since people are assumed to be arriving at the bus stop uniformly over time and one person boards/alights at a time, all we need to do is to identify the luckiest and unluckiest people \emph{with respect to each bus}, viz. the one who arrives at the bus stop just when the bus is there and about to leave \emph{or} the last person to be allowed boarding before the no-boarding policy is implemented (hence would experience the shortest waiting time), as well as the one who arrives at the bus stop right after the previous bus had just left the bus stop \emph{or} was just being denied boarding by the previous bus (hence would experience the longest waiting time). Once we know the waiting times for the luckiest and unluckiest people, the average is just half of the sum of their waiting times. Also as mentioned earlier, in steady state, each bus picks up the same number of people (on average). So bus B, being ``slower'', would pick up $L$ people (on average), and then implement no-boarding and leave, leaving the rest of the people to wait for bus A --- who would also pick up a total of $L$ people (on average). Let us now look at each bus:
\begin{enumerate}
\item For waiting to board the ``slower'' bus B, the unluckiest person is the one who is at the bus stop when bus A just left. Since the effective phase difference is $\theta_\textrm{eff}$, this person needs to wait for $\theta_\textrm{eff}/\omega=\theta_\textrm{eff}T/2\pi$ before bus B arrives, and a further $\tau/2$ for passengers alighting before this person could board.

The so-called luckiest person to board bus B is the last person allowed to board before the no-boarding policy is implemented. This last person would have waited $\theta_\textrm{eff}/\omega-(T+\tau)/2$ for bus B to arrive. This duration is $(T+\tau)/2$ less than the unluckiest person because bus B would pick up half of the total number of people per revolution [which is half of $s(T+\tau)$ on average], who would have accumulated at the bus stop over a time period of $(T+\tau)/2$ since bus A had left. After bus B arrives, this person has to wait a further $\tau/2$ for people to alight with another $\tau/2$ for those ahead to board first, before finally boarding, i.e. a total of $\theta_\textrm{eff}T/2\pi-T/2+\tau/2$.

(From here, we can see that the person who is just denied boarding by bus B would have waited $\theta_\textrm{eff}/\omega-T/2-\tau/2$, and then wait for another $T-\theta_\textrm{eff}/\omega$ for bus A to arrive --- which is a total of $T/2-\tau/2$. This will be used below.)

Hence on average, the waiting time to board bus B is
\begin{align}
W_B&=\frac{1}{2}\left(\left(\frac{\theta_\textrm{eff}T}{2\pi}+\frac{1}{2}\tau\right)+\left(\frac{\theta_\textrm{eff}T}{2\pi}-\frac{1}{2}T+\frac{1}{2}\tau\right)\right)\\
&=\frac{\theta_\textrm{eff}T}{2\pi}-\frac{1}{4}T+\frac{1}{2}\tau.
\end{align}
\item For waiting to board the ``faster'' bus A, the unluckiest person who was just denied boarding by bus B has to wait a total of $T/2-\tau/2$ for bus A to eventually arrive (as mentioned above), plus $\tau/2$ for people to alight before this person could board. In contrast, the luckiest person who just arrives when bus A is at the bus stop and about to leave would have zero waiting time. Thus on average, the waiting time to board bus A is
\begin{align}
W_A&=\frac{1}{2}\left(\left(\frac{1}{2}T\right) + (0)\right)\\
&=\frac{1}{4}T.
\end{align}
\end{enumerate}
Since bus A and bus B pick up, on average, an equal number of passengers each time, the overall average waiting time for this system is 
\begin{align}
W&=\frac{1}{2}\left(W_A+W_B\right)\\
&=\frac{1}{2}T\left(\frac{\theta_\textrm{eff}}{2\pi}\right)+\frac{1}{4}\tau\\
W(x, k, T)&=\frac{1}{2}Tx+\frac{1}{4}\tau\\
\overline{W}(x, k)&=\frac{1}{2}x+\frac{1}{4}\overline{\tau}(k),\label{waiting}
\end{align}
In the third line, we normalise the effective phase difference between the two buses by $x:=\theta_\textrm{eff}/2\pi$ with $x\in[x_\textrm{min}, 1]$, $x_\textrm{min}=\theta_\textrm{min}/2\pi>1/2$ [Eq.\ (\ref{lowerbound})]. Note that Eq.\ (\ref{lowerbound}) implies that whilst the ideal situation is the antipodally staggered configuration where $\theta_\textrm{eff}=\pi$ or $x=1/2$, the lower bound to implement no-boarding is necessarily deviated due to $\overline{\tau}$, as the bus has to stop for alighting/boarding. Like $\overline{\tau}$, we have defined $\overline{W}$ as the waiting time per unit $T$. In conclusion, the no-boarding policy linearly reduces the average time for passengers waiting to board the bus with respect to the effective phase difference between these two buses. The implemented angle $\theta_0$ must be such that the effective angle $\theta_\textrm{eff}$ is not smaller than $\theta_\textrm{min}$ in Eq.\ (\ref{lowerbound}) otherwise the bus system is not picking up people at a rate higher than the rate of people demanding service.

\subsection{Simulations based on real bus loop service data}\label{NTUsimulations}

A recent study on Singapore's Nanyang Technological University's (NTU) campus shuttle bus system comprising $N$ buses serving $M=12$ reasonably staggered bus stops in a loop was carried out to investigate the mechanisms of bus bunching \cite{Vee2019}. The study reported that these buses typically have different natural speeds, due to different human drivers' driving styles. There are distinctively slow drivers and fast drivers. Their average periods around the loop without stopping are within a range of 12 to 18 minutes, i.e. $f_i\in[0.93, 1.39]$ mHz. Further data analysis on this shuttle bus system found that the ratio of the rate of people arriving at each bus stop to the rate of people loading up the bus, $k:=s/l$, averaged over all $M=12$ bus stops, is of the order of $k\sim0.020$ during lull periods and $k\sim0.065$ during busy periods. This information implies that with the people taking about a second to get up the bus so that $l\sim1$ person per second, then the arrival rate of people at a bus stop is $s\sim0.020$ person per second ($1.2$ person per minute) during lull periods and $s\sim0.065$ person per second (almost $4$ people per minute) during busy periods.

Based on these real data for $M=12$ bus stops, we can test our analytical theory of the no-boarding policy for a realistic simulated system having $N=2$ buses serving $M=1$ bus stop in a loop. Let the two buses have identical natural period of $T=12$ minutes to complete one loop without stopping. Identical natural period would be applicable to programmable self-driving buses which would be rolled out in the near future \cite{NTUautobuses}. The case of human-driven buses with different natural periods is investigated in Section 5. Normally, there are three or four buses serving the route, with up to seven or even eight buses during busy times. Since we are only dealing with two buses here, we consider a lull period with $k=0.010$. This would then produce reasonable number of people demanding service and they are able to fit within the buses' usual capacity of an order of $50$ passengers. Nevertheless, we emphasise here that these results hold for any value of $k$ even in busy periods, since there is nothing in the theory that specifically restricts to the lull period or any value of $k$. Our use of a lull $k=0.010$ in this subsection is simply because we are here considering $N=2$ buses and pushing up $k$ would make the number of people on each bus higher --- which is fine, unless a maximum bus capacity is additionally imposed. Subsequently when we generalise to $N$ buses (last part of this section, below), then more people (higher $k$, busy period) can be served without unrealistically carrying ``hundreds of people per bus''.

Anyway, suppose each person takes $l=1$ second to board or alight, which translates to a people arrival rate of $s=0.010$ person arriving per second or one person arrives every 100 seconds. Since $k=0.010$ is a value applicable to $M=12$ bus stops in the NTU bus system, for our simulations with only $M=1$ bus stop, we should multiply this arrival rate by $12$ to generate a comparable proportion of people using the service, i.e. a person arriving every $100/12$ seconds. Note that typically people would travel from one part of the loop to another, travelling say an average of about half a loop (or less, since people can take the bus service going in the opposite direction for a shorter path), rather than making one full loop. As our simulations with $M=1$ bus stop force people to travel one full loop, we halve the arrival rate in order to more realistically reflect the number of people on the two buses. Therefore, we have one person arriving every $100/6$ seconds or rounded down to $16$ seconds (because the simulations run on discrete time steps of $1$ second).

In the usual situation with no intervention which corresponds to setting $\theta_0=360^\circ=2\pi$ radians, the two buses bunch and would remain so since their natural speeds are identical. In steady state, our simulation results are as follows: The average waiting time for passengers is $0.515\pm0.299$ units of $T=12$ minutes ($0.299$ is one standard deviation, since people arrive uniformly over time with some having to wait longer than others for the bus to arrive), the time spent on the bus is $1.032$ units (there is no uncertainty here, since every passenger spends this same amount of time on the bus), giving a total travel time of $1.546\pm0.299$ units. Both buses spend $\overline{\tau}=0.067$ units stopping at the bus stop (there is no uncertainty here, since the bunched pair of buses always spend this same amount of time stopping at the bus stop) to allow alighting followed by boarding (bunched buses share the load). Each bus load carries $L=24$ people. These value of $\overline{\tau}$ and $L$ are exactly the values predicted by Eqs.\ (\ref{tau})-(\ref{people}). In addition, Eq.\ (\ref{waiting}) gives for $\theta_\textrm{eff}=2\pi$ radians ($x=1$) and $k=1/16$, an average waiting time of $\overline{W}=0.517$ units which is in agreement with our simulation value.

\begin{figure}
\centering
\includegraphics[width=12cm]{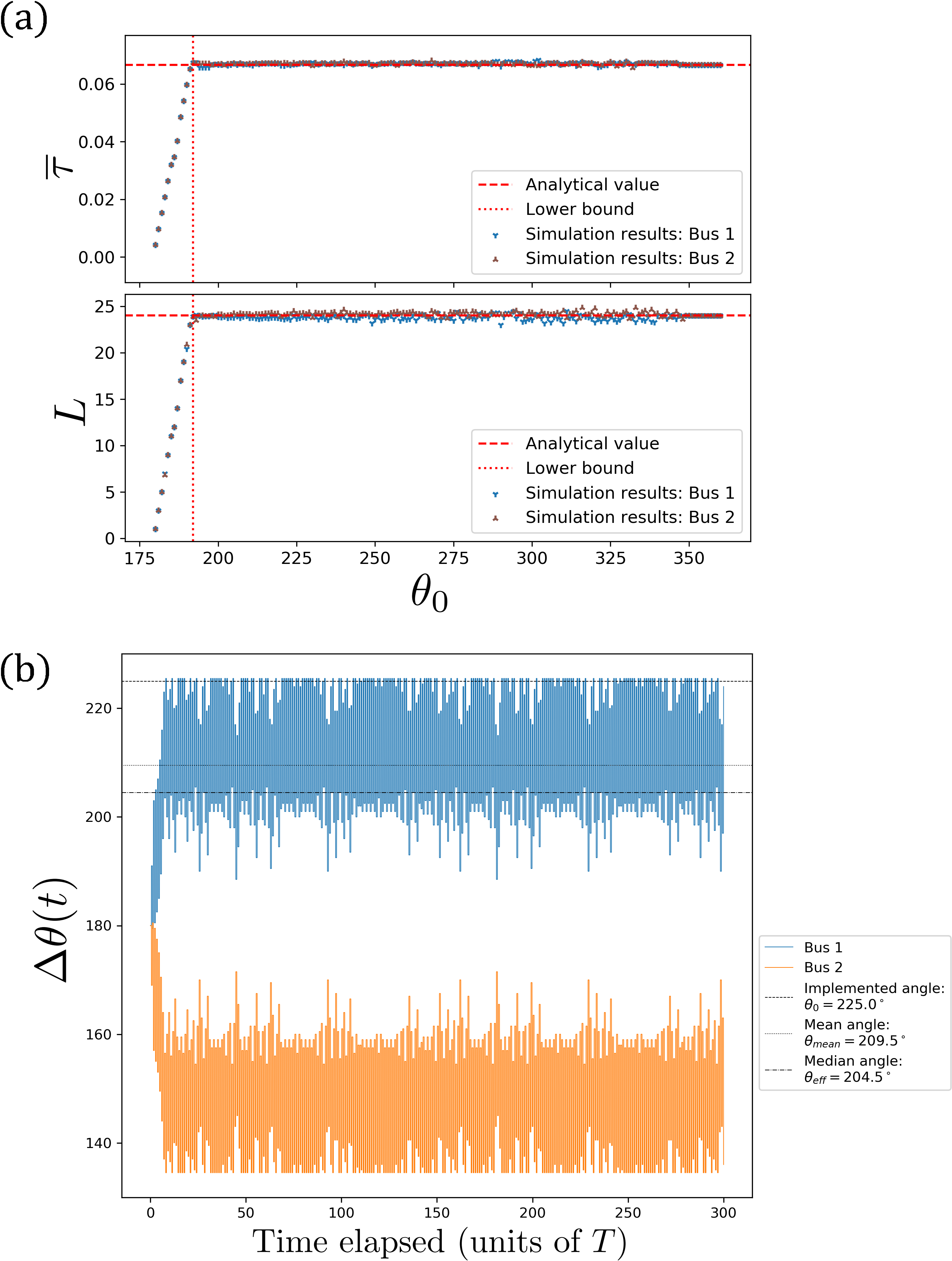}
\caption{(a) Graphs of $\overline{\tau}$ and $L$ versus $\theta_0$ for each of the two buses. These simulation results agree with Eqs.\ (\ref{tau})-(\ref{people}), as well as the lower bound $\theta_\textrm{min}$ from Eq.\ (\ref{lowerbound}). Shown here [as well as in (b)] is for $k=1/16$, and these features similarly hold for any value of (fixed) $k$. (b) Graphs of $\Delta\theta(t)$ versus time for each of the two buses. The no-boarding policy kicks in if $\Delta\theta>225^\circ$.}
\label{fig3}
\end{figure}

Next, we show results for this lull period where a person arrives every $16$ seconds at this single bus stop, with the no-boarding policy in action. We carry out separate simulations for various (but fixed) $\theta_0\in\{180^\circ, 181^\circ,\cdots,360^\circ\}$ ($360^\circ$ means no implementation). Fig.\ \ref{fig3}(a) indicates excellent agreement between the analytical theory and the simulation results (which are averaged over many rounds during steady state). The stoppage duration is typically averaged around $\overline{\tau}=0.067$ units for both buses [with standard deviation of the order of $0$ to $0.025$ units, depending on the actual $\theta_0$ being implemented, since $\Delta\theta(t)$ fluctuates around $\theta_\textrm{eff}$ and depends on $\theta_0$], and the number of people that they pick up is about $L=24$ each (with standard deviation of the order of $0$ to $20$ people, depending on $\theta_0)$. These graphs also affirm that when the implemented angle is below the lower bound $\theta_0<\theta_\textrm{min}$, then $\tau$ and consequently $L$ are less than the demand level [see also Fig.\ \ref{fig5}, which shows the lower bound $\theta_\textrm{min}$ for any $k$].

Fig.\ \ref{fig3}(b) shows the phase difference for one of these runs between the two buses, measured from each with respect to the other (so they always sum up to $360^\circ$) as a function of time. They start off antipodally staggered and very quickly one bus ``goes chasing after'' the other ``slower one''. Here, the no-boarding policy is implemented if $\Delta\theta>\theta_0=225^\circ$ which prevents them from bunching, and they settle on a steady state where their phase difference fluctuates around the median of $204.5^\circ$ or mean of $209.5^\circ$. From the simulations, the average waiting time is $0.294\pm0.163$ units, certainly much shorter than the bunched case of $0.515$ units. It turns out that according to Eq.\ (\ref{waiting}), it is the median phase difference of $\theta_\textrm{eff}=204.5^\circ$ corresponding to an average waiting time of $\overline{W}=0.301$ units, which closely matches the simulation value. In contrast, the mean phase difference plugged into Eq.\ (\ref{waiting}) gives $0.308$ units, which is a slight overestimate.

\begin{figure}
\centering
\includegraphics[width=15cm]{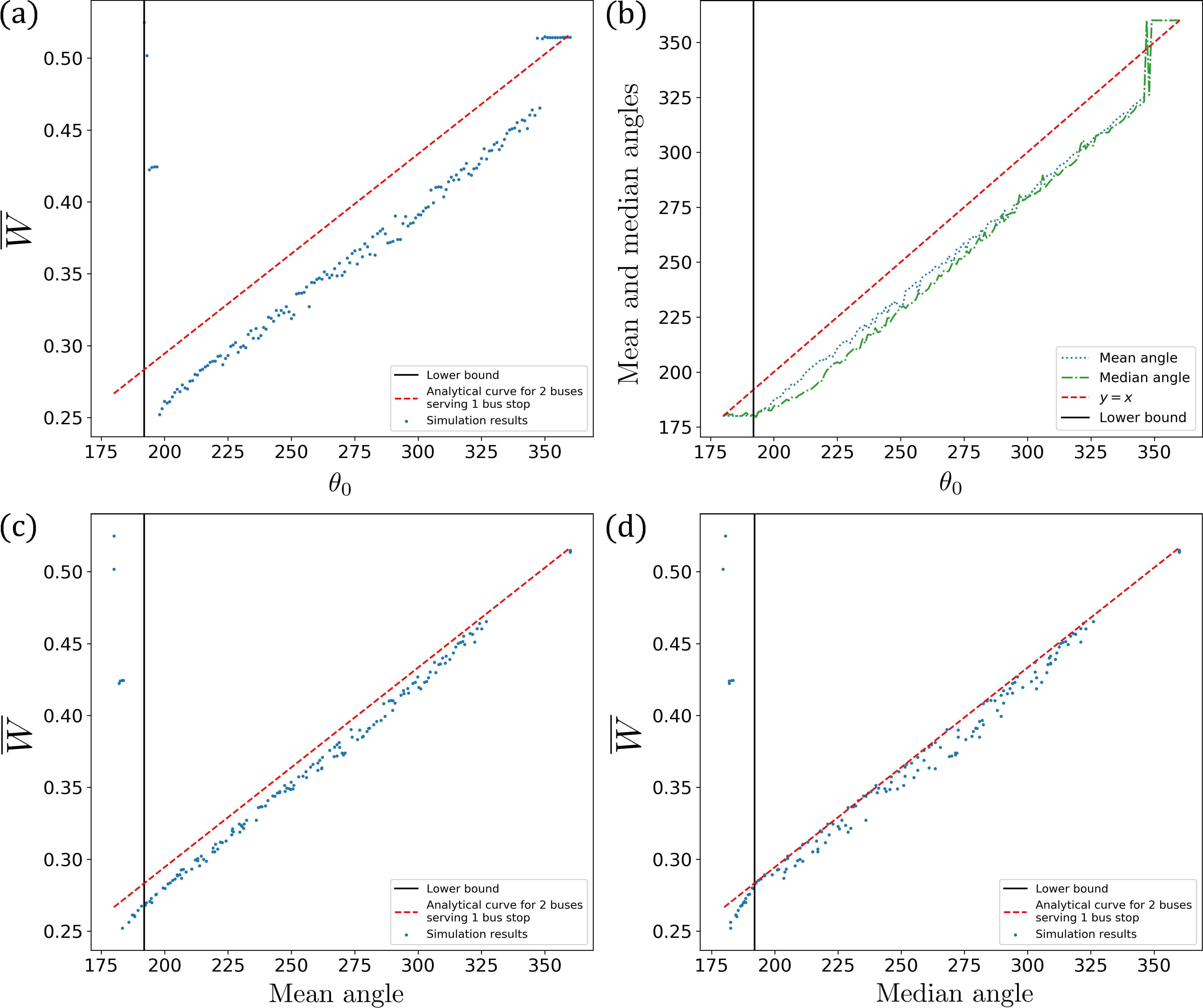}
\caption{Graphs of $\overline{W}$ versus: (a) $\theta_0$, (c) the mean, as well as (d) median phase differences, respectively. The effective phase difference $\theta_\textrm{eff}$ during steady state being the \emph{median phase difference} between the two buses more closely matches the predicted value from Eq.\ (\ref{waiting}), compared to the mean phase difference. Note that below the lower bound $\theta_0=\theta_\textrm{min}$, $\overline{W}$ blows up by two orders of magnitude (not plotted): $\overline{W}=10.4$ at $\theta_0=191^\circ$, $\overline{W}=31.5$ at $\theta_0=190^\circ$, $\overline{W}=54.6$ at $\theta_0=189^\circ, \cdots$. Similar graphs can be obtained for any value of $k$. (b) The relationship between $\theta_0$ and the mean as well as median phase differences as measured from the simulations.}
\label{fig4}
\end{figure}

These all generalise to various implemented angles $\theta_0$ as shown in Fig.\ \ref{fig4}. Shown here is a collection where each simulation data point is an independent run with different (but fixed) $\theta_0$. In each plot of Figs.\ \ref{fig4}(a), (c) and (d), the (red) dash line is the analytical result from Eq.\ (\ref{waiting}), whilst the (blue) dots are simulations points. Fig.\ \ref{fig4}(a) shows a ``na\"{i}ve'' plot of the simulation results for the average waiting time versus the implemented angle $\theta_0$. There is some conspicuous shift between the analytical line and the simulation points. This is because the effective phase difference $\theta_\textrm{eff}$ is smaller than the implemented one $\theta_0$, with the exception of the ``horizontal tail'' of points for $\theta_0\sim360^\circ$. That horizontal tail appears because for those extremely large $\theta_0$, the no-boarding policy is unable to prevent bunching and once bunched the buses stay bunched, i.e. their effective phase difference is $360^\circ$. But what is a suitable representative value of $\theta_\textrm{eff}<\theta_0$ in general? Fig.\ \ref{fig4}(b) plots the relationship between $\theta_0$ and the mean as well as median angles, as measured from the steady state part of the simulations. The mean seems to be biased, in the sense that the gradient of the ``mean line of points'' is not parallel to $y=x$, i.e. as the implemented angle $\theta_0$ gets smaller, the degree of shift on the mean angle is smaller. On the other hand, the median angle shows a more consistent shift across various implemented angles $\theta_0$. Consequently in Fig.\ \ref{fig4}(c), when the simulation data points are adjusted based on the mean phase difference, there is still a noticeable deviation from the analytical line. Nevertheless, this difference is not too glaring when adjusted based on the median phase difference in Fig.\ \ref{fig4}(d).

\begin{figure}
\centering
\includegraphics[width=15cm]{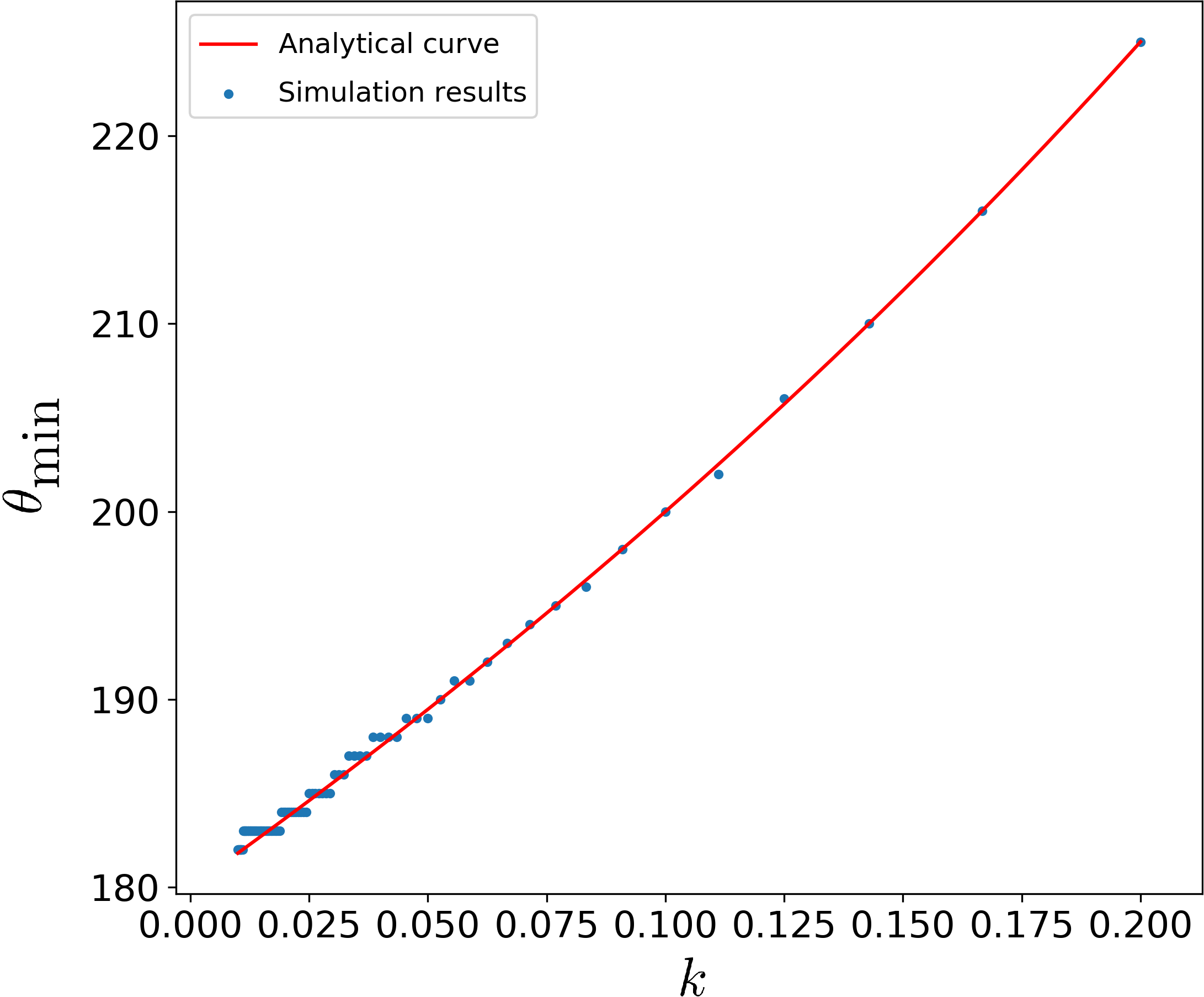}
\caption{Graph of $\theta_\textrm{min}$ versus $k$.}
\label{fig5}
\end{figure}

Furthermore, we run separate simulations for this bus system where the no-boarding policy is implemented at various (but fixed) angles $\theta_0\in\{180^\circ, 181^\circ, \cdots, 360^\circ\}$, for various (but fixed) values of $k\in\{1/100, 1/99, \cdots, 1/5\}$. With these extensive simulation data, we find that the analytical theory's predictions are in excellent agreement with the simulation results. In particular, it holds for any (fixed) $k$ which comprises both lull and busy periods. In fact, the analytical curve for $\theta_\textrm{min}(k)$ in Eq.\ (\ref{lowerbound}) fits nicely as displayed in Fig.\ \ref{fig5}, where simulations show that if $\theta_0<\theta_\textrm{min}$, then the waiting times for the passengers blow up to multiple (tens of) revolutions. As explained, this is because the buses are disallowing boarding \emph{too frequently} whenever their phase difference deviates too much from the staggered configuration and $\Delta\theta>\theta_0$ executes no-boarding, to the point where the bus system is not meeting the demand for service. The number of people at the bus stop would grow unbounded as time goes to infinity, when $\theta_0<\theta_\textrm{min}$. If $\theta_0>\theta_\textrm{min}$, then this number remains bounded although it may not always drop to zero since the no-boarding policy would leave some people there.

\subsection{\texorpdfstring{$N=2$}{N=2} buses serving many bus stops}

With more than $M=1$ bus stops, each additional bus stop systematically contributes towards an additional stoppage due to $\tau$ where the bus allows alighting/boarding. Obviously such intermediate stoppages at each such bus stop in between would bump up the waiting time. The overall trend corresponds to the analytical result for $N=2$ buses serving $M=1$ bus stop [see Fig.\ \ref{fig6}(b), below]: the average waiting time decreases linearly with a decrease in $\theta_0$ where no-boarding is implemented if $\Delta\theta>\theta_0$. In particular, the general gradient of the simulation points matches the analytical line but is shifted upwards due to additional stoppages at other bus stops. On top of that, in Fig.\ \ref{fig6}(b) with $M=12$ staggered bus stops being $30^\circ$ apart from each other, we observe discrete jumps at $210^\circ, 240^\circ, 270^\circ, 300^\circ, 330^\circ$, where these additional bus stops are located between the two buses.

\subsection{\texorpdfstring{$N$}{N} buses serving \texorpdfstring{$M$}{M} bus stops --- analytical theory and simulations}

We can directly generalise the analytical theory from $N=2$ buses to any $N\geq2$ buses serving $M=1$ bus stop. Here are the results:
\begin{enumerate}
\item The stoppage duration for each bus is
\begin{align}\label{tauN}
\overline{\tau}(k,N):=\frac{\tau}{T}=\frac{2k}{N-2k}=\sum_{n=1}^\infty{\left(\frac{2k}{N}\right)^n}.
\end{align}
The geometric series expansion is valid, since $k<1$ and $N\geq2$.
\item The lower bound to $\theta_0$ is
\begin{align}\label{generalLower}
x_\textrm{min}(k,N):=\frac{\theta_\text{min}}{2\pi}=\frac{1}{N}\left(1+\overline{\tau}(k,N)\right).
\end{align}
\item The average waiting time for passengers at the bus stop before a bus arrives is as follows. For $N$ buses, there are $N-1$ piecewise continuous line segments. The $i$-th line segment, where $i=1,2,\cdots,N-1$, is given by
\begin{align}\label{generalW}
\overline{W}(x,k,N):=\frac{W}{T}=\frac{i(i+1)}{2N}x+\frac{1}{2}-\frac{i}{N}+\frac{1}{4}\overline{\tau}(k,N),
\end{align}
where $1/(i+1)\leq x\leq1/i$ and $x:=\theta_\textrm{eff}/2\pi$.
\end{enumerate}
Eq.\ (\ref{tauN}) is derived by having a total of $s(T+\tau)$ people per revolution being carried by the $N$ buses, so each bus carries $s(T+\tau)/N$ and they do so over an average time interval of $\tau/2$ (since the other $\tau/2$ is for people to alight). This gives $s(T+\tau)/N=l\tau/2$, which leads to Eq.\ (\ref{tauN}). For Eqs.\ (\ref{generalLower}) and (\ref{generalW}) though, we directly calculated all these expressions individually and case-by-case by hand up to $N=6$ in the manner similar to the $N=2$ buses serving $M=1$ bus stop case in Section 3A, and then write down the generalisation to any arbitrary $N=2,3,\cdots$ buses serving $M=1$ bus stop. In particular for Eq.\ (\ref{generalW}) the calculation steps are exactly those for $N=2$ buses, which are applied systematically and tediously to more and more buses, one by one, calculating the luckiest and unluckiest people to board each bus, taking note of the proportion of load carried by each bus or each group of bunched buses.

Let us illuminate the content of Eq.\ (\ref{generalW}). For $N$ buses, there are $N-1$ piecewise continuous line segments labeled by $i=N-1,N-2,\cdots,1$, corresponding to increasing $x$ from $x_\textrm{min}$ to $1$. The gradient of the line segment labelled by $i=N-1$ is the steepest, and these line segments monotonically become less steep as we decrease $i$ until $i=1$ (but all have positive slope). The line $i=N-1$ is the situation where all $N$ buses are unbunched. For $i=N-2$, there is a pair of bunched buses, with the remaining all unbunched. For $i=N-3$, there is a triad of three buses bunched into one unit, with the remaining all unbunched. This goes on until $i=1$, where all $N$ buses bunch into one single unit.

\begin{figure}
\centering
\includegraphics[width=15cm]{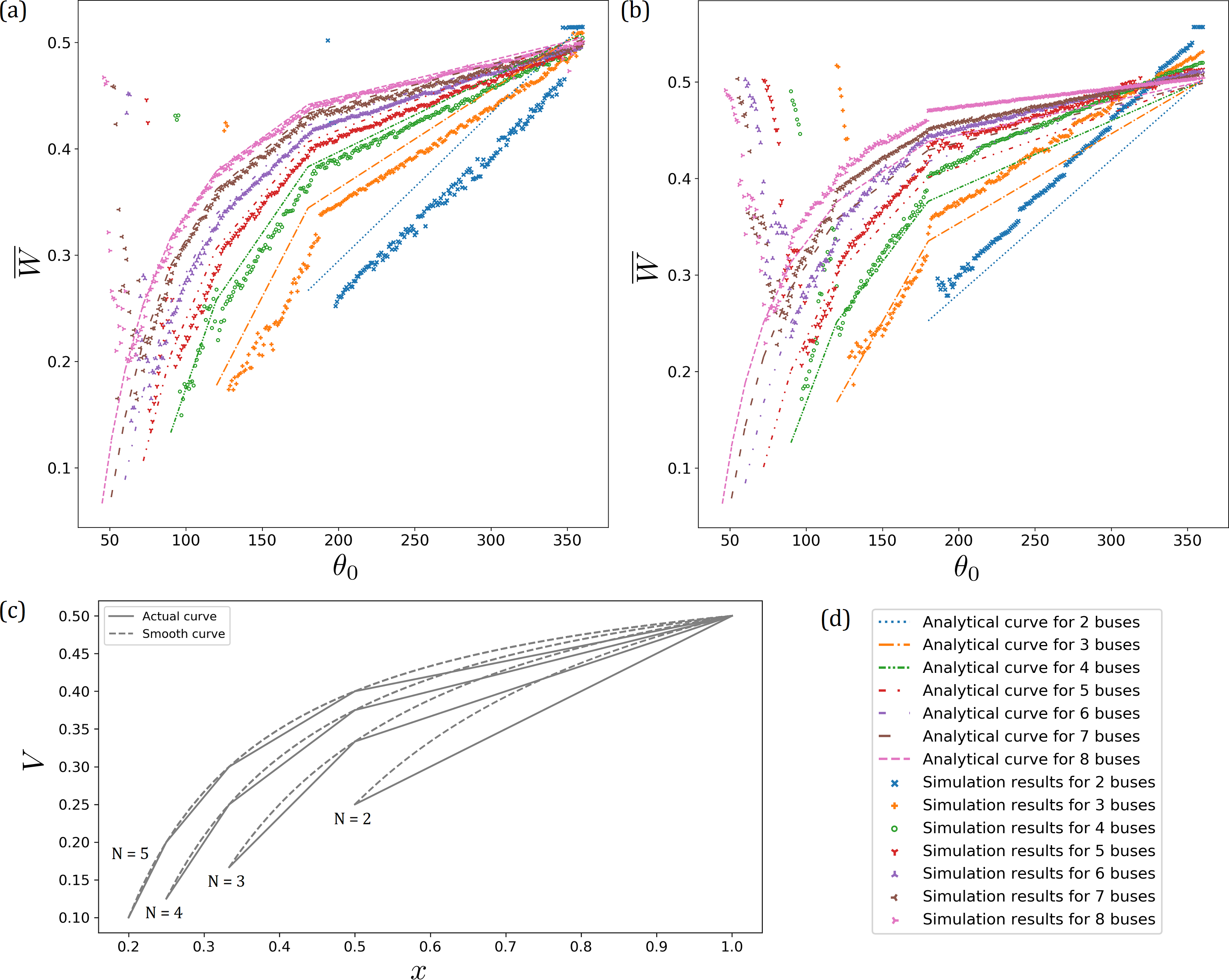}
\caption{(a) Simulation results for $N=2,3,4,5,6,7,8$ buses, respectively, serving $M=1$ bus stop. The demand level is fixed at $k=1/16$. (b) Simulation results for $N=2,3,4,5,6,7,8$ buses, respectively, serving $M=12$ bus stops. The demand level is fixed at $k=1/100$. The analytical curves are for those serving $M=1$ bus stop. The simulation points have average waiting times which are higher than their corresponding analytical curves, since these buses serve $M=12$ bus stops. Furthermore, there are discrete jumps on the simulation points whenever an extra bus stop is present. (c) Graphs of the piecewise continuous curves for $N=2,3,4,5$ as given by Eq.\ (\ref{generalW}) with $\overline{\tau}/4$ subtracted away, together with the corresponding smooth curves Eq.\ (\ref{curve}) that pass through the boundary points of each line segment, for the respective $N$. These smooth curves are \emph{hyperbolas}, and approach an ``$L$''-shaped asymptotic curve as $N\rightarrow\infty$. (d) Legend for (a) and (b).}
\label{fig6}
\end{figure}

Fig. \ref{fig6}(a) shows the simulation results for $N=2,3,4,5,6,7,8$, respectively, serving $M=1$ bus stop in a loop, with the corresponding analytical curves Eq.\ (\ref{generalW}). Note that the gap between the analytical curve and the simulation points is due to this plot being with respect to the implemented angle $\theta_0$ instead of the effective angle $\theta_\textrm{eff}$. With $N$ buses, there are $N-1$ independent local phase differences between adjacent buses, since these $N$ local phase differences sum up to $2\pi$. Eq.\ (\ref{generalW}) with $N>2$ assumes that all $N-1$ independent effective angles between any local pair of adjacent buses are equal, whereas in the actual system, they are not and sometimes even differ substantially. Making a single representative $\theta_\textrm{eff}$ out of those different local phase differences would introduce bias, sometimes giving grossly wild results especially with larger $N$. Plotting against $\theta_0$ would cleanly circumnavigate this and not artificially tamper with the data points, though we would have that shift between the analytical curve and the simulation points since $\theta_\textrm{eff}<\theta_0$.

As mentioned, each of these line segments in Eq.\ (\ref{generalW}) represents different situations corresponding to different number of bunched buses. For example with $N=3$, when the implemented angle for no-boarding $\theta_0$ is such that $\theta_\textrm{min}<\theta_\textrm{eff}<180^\circ$, the three buses are not bunched. However, if $\theta_0$ is larger such that $180^\circ<\theta_\textrm{eff}<360^\circ$, then two of these buses would bunch such that the system becomes a two-bus system where one of them comprises two bunched buses with twice the loading rate. It turns out that this is actually \emph{less efficient} than a system with two unbunched buses where they have the same loading rate. Intuitively, this is because in the former the bunched pair is so-called ``faster'' due to its double loading rate and chases after the lone ``slower'' bus which has to implement no-boarding. The lone bus only picks up $1/3$ of the demand, whilst the bunched pair picks up $2/3$ of the demand ($1/3$ each). Similar ramifications are true for $N=4,5,6,7,8$ buses, as shown in Fig.\ \ref{fig6}(a). The bunched pair becomes ``faster'' and grows into larger bunched group of buses, if the implemented angle $\theta_0$ is large. A bus system with larger bunched group of buses performs less efficiently compared to smaller bunched group of buses.

With many bus stops, each bus stop adds additional waiting time, just like the case of $N=2$ buses discussed in the preceding subsection. Fig.\ \ref{fig6}(b) shows simulation results for $N=2,3,4,5,6,7,8$ buses, respectively, serving $M=12$ bus stops. In these simulations, we set $k=0.010$. Nevertheless, similar results hold for any (fixed) value of $k$. As the number of buses increases, since we keep the demand level to be the same lull $k=0.010$, having more buses would spread out the load and so each bus would spend shorter $\overline{\tau}$ at the bus stops [Eq.\ (\ref{tauN})]. Therefore, the jump at each bus stop diminishes with larger $N$. 

We do not proceed to repeat the laborious calculations for the general case of $N$ buses serving $M$ bus stops, because the analytical results from $N$ buses serving $1$ bus stop are sufficient to describe and understand the behaviour as $\theta_0$ is changed. As the simulation results indicate in Fig.\ \ref{fig6}(b), the precise corrections from what would be painstakingly going through each piece of individual case-by-case calculations only amounts to adjusting for the small shifts due to additional bus stops without affecting the already known behaviour of the system.

There is one important remark regarding decreasing the implemented angle $\theta_0$ towards $\theta_\textrm{min}$. For each $N$, as $\theta_0\rightarrow\theta_\textrm{min}^+$, after the linear decrease, the average waiting time begins to increase before eventually blowing up by several orders once $\theta_0<\theta_\textrm{min}$ [see Fig.\ \ref{fig6}(a) and (b)]. These increases in the average waiting time when $\theta_0\rightarrow\theta_\textrm{min}^+$ are more prominent with more buses, suggesting that with more buses the system is more susceptible to perturbations which nudge the buses to implement no-boarding too soon as its $\Delta\theta(t)$ gets larger than $\theta_0$. In particular, adding more buses does not always seem to lead to the expected improved average waiting time near $\theta_\textrm{min}$. For instance in Fig.\ \ref{fig6}(b), the simulated best average waiting time for $N=8$ buses is higher than $0.2$ units, way above its theoretical minimum of less than $0.1$ units. In contrast, the $N=3$ and $N=4$ systems do achieve simulated best average waiting times below $0.2$ units, which are better than the $N=8$ system. Therefore, whilst it is desirable to implement $\theta_0$ as small as possible, one needs to beware that setting it too close to $\theta_\textrm{min}$ would risk having the system implement no-boarding too frequently due to fluctuations in $\Delta\theta(t)$.

Incidentally, note that in the expression for the average waiting time in Eq.\ (\ref{generalW}), the $k$-dependence solely arises from $\overline{\tau}$. If we define the quantity $V(x,N):=\overline{W}(x,k,N)-\overline{\tau}(k,N)/4$, then this quantity is purely a function of $x\in[x_\textrm{min},1]$, for each $N$. We can immediately write down some properties of $V(x,N)$ [see also Fig.\ \ref{fig6}(c)]:
\begin{enumerate}
\item The set of boundary points for each of the $N-1$ line segment constituents for each curve of constant $N$ is
\begin{align}\label{boundarypoints}
\left\{\left(\frac{1}{i},\frac{1}{2}+\frac{1}{2N}(1-i)\right), i=1,2,\cdots,N\right\}.
\end{align}
The leftmost boundary point is when $i=N$, i.e.
\begin{align}
\left(\frac{1}{N},\frac{1}{2N}\right),
\end{align}
whilst the rightmost boundary point is when $i=1$, i.e.
\begin{align}
\left(1,\frac{1}{2}\right).
\end{align}
The rightmost boundary point for any curve of constant $N$ is the same point. This is where all buses bunch into a single unit, and the average waiting time is half units of $T$ (plus $\overline{\tau}/4$).
\item The curve
\begin{align}\label{curve}
V(x)=\frac{1}{2}+\frac{1}{2N}\left(1-\frac{1}{x}\right)
\end{align}
passes through all those points in Eq.\ (\ref{boundarypoints}). This equation shows that the \emph{overall} decrease in average waiting time as $x$ decreases (or as $\theta_0$ approaches the \emph{staggered configuration}) is in fact, \emph{hyperbolic}.
\item With Eq.\ (\ref{curve}), it is clear that as $N\rightarrow\infty$, the curves shift upwards and leftwards, approaching an ``$L$''-shaped asymptotic curve defined by the horizontal line $y=1/2$ (from $x=0$ to $x=1$) and the vertical line $x=0$ (from $y=0$ to $y=1/2$). This is a manifestation of a larger group of bunched buses being less efficient than a smaller bunched group due to greater asymmetry in the system. Higher $N$ also allows for further reduction in waiting time, since there are more buses --- with diminishing returns, however.
\item The set of leftmost boundary points for each curve of constant $N$, viz.
\begin{align}
\left\{\left(\frac{1}{N},\frac{1}{2N}\right), N=2,3,4,\cdots\right\},
\end{align}
as well as their common rightmost boundary point $(1,1/2)$, all lie on the straight line $y=x/2$.
\item Eq.\ (\ref{generalW}) readily extends to include the degenerate situation of $N=1$ bus. Here, $\theta_\textrm{eff}=2\pi$ as measured from itself to itself since there is just one bus serving the loop, and the average waiting time is always half units plus $\overline{\tau}/4$. So Eq.\ (\ref{generalW}) is just the point $(1,1/2+\overline{\tau}/4)$ for $N=1$, with $i=0,x=1$ (or $\theta_\textrm{eff}=2\pi$), $\overline{\tau}=2k/(1-2k)$. In this case, $k$ has to be less than $1/2$ in order for that lone bus to meet the demand for service.
\end{enumerate}

\section{No-boarding policy: Looking behind}

Looking at the bus \emph{immediately ahead} seems natural, since the number of people to be picked up is directly determined by how long ago the bus immediately ahead has left. But is this the only way to decide on a no-boarding policy? Instead of looking at the bus immediately ahead, it is also possible to consider measuring the phase difference with respect to the bus \emph{immediately behind} \cite{Daganzo09,Bart12}. When a bus is boarding people at a bus stop, if the phase difference with respect to the bus immediately behind $\Delta\theta$ is less than some threshold $\theta_0$, it is an indication that it is ``too slow'' and bunching may be imminent. To prevent this, it should disallow further boarding and leave. Similar calculations can be carried out, analogous to the looking-ahead rule in Section 3. Firstly, the average stoppage time spent by each of the $N$ buses serving $1$ bus stop in a loop is the same as in Section 3, given by Eq.\ (\ref{tauN}). Therefore, the average number of people picked up is also the same as in Section 3.

The threshold $\theta_0$ for implementation of no boarding must be less than $2\pi/N$ in a system with $N$ buses. For example let $N=2$ and set $\theta_0=181^\circ$, where no boarding is implemented if $\Delta\theta<181^\circ$. Consider the staggered configuration where the two buses have a phase difference of $180^\circ$. According to this threshold $\theta_0=181^\circ$, whenever one bus is at a bus stop, it measures $\Delta\theta=180^\circ$ which is less than $\theta_0=181^\circ$ and implements no boarding. This would always be the case, and nobody would ever be picked up! Hence in general, one should choose $\theta_0<2\pi/N$. For $N=2$, we can calculate this upper bound more precisely to be
\begin{align}\label{strictupperbound}
x_\textrm{max}(k,N):=\frac{\theta_\text{max}}{2\pi}=\frac{1}{2}\left(1-\overline{\tau}(k,N)\right).
\end{align}
The derivation is analogous to calculating the lower bound to $\theta_0$ in the looking-ahead rule in Section 3. For $N>2$ however, the corresponding method does not work since looking behind is not equivalent to looking ahead.

\begin{figure}
\centering
\includegraphics[width=16cm]{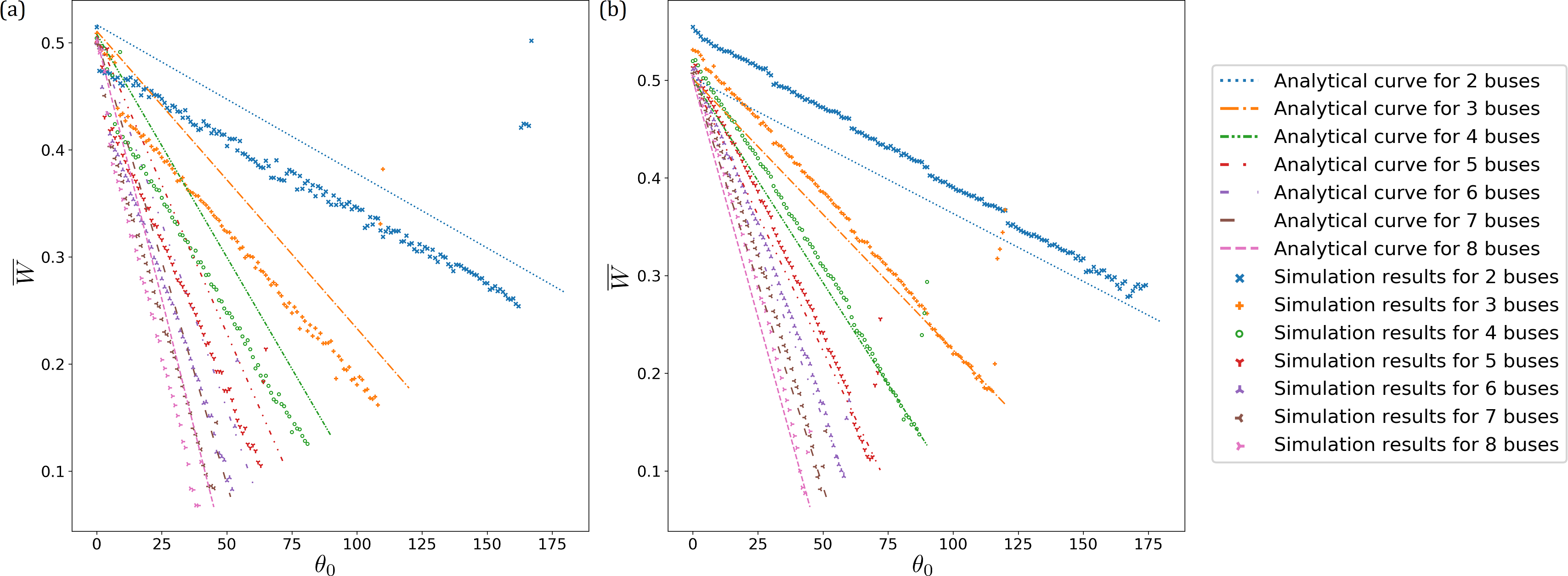}
\caption{(a) Simulation results for $N=2,3,4,5,6,7,8$ buses, respectively, serving $M=1$ bus stop. The demand level is fixed at $k=1/16$. (b) Simulation results for $N=2,3,4,5,6,7,8$ buses, respectively, serving $M=12$ bus stops. The demand level is fixed at $k=1/100$. The analytical curves are those serving $M=1$ bus stop. The simulation points have average waiting times which are higher than their corresponding analytical curves, since these buses serve $M=12$ bus stops. Furthermore, there are discrete jumps on the simulation points whenever an extra bus stop is present.}
\label{fig7}
\end{figure}

Recall from the previous section on looking ahead where increasing $\theta_0$ from $2\pi/N$ to $2\pi$ would lead to the $N$ buses undergoing transitions where buses progressively bunch into one large group. This is different for looking behind. For all values of $\theta_0\in(0,2\pi/N)$, the $N$ buses are always in a completely unbunched configuration (and increasing $\theta_0$ beyond $2\pi/N$ would lead to the buses always implementing no boarding, as explained above). The analytical results for the average waiting time for a passenger waiting at a bus stop for a bus to arrive, in a system of $N$ buses serving $1$ bus stop is:
\begin{align}
\overline{W}(x,k,N):=\frac{W}{T}=-\left(\frac{N-1}{2}\right)x+\frac{1}{2}+\frac{1}{4}\overline{\tau}(k,N),
\end{align}
where $x:=\theta_\textrm{eff}/2\pi$. This relation between $\overline{W}$ and $x$ is the same as that of Eq.\ (\ref{generalW}) with $i=N-1$ (which corresponds to $N$ unbunched buses), i.e. it has the same gradient in terms of magnitude, and peaks with a value of $1/2+\overline{\tau}/4$ (when $x=0$) where all buses bunch into one single unit. [In Eq.\ (\ref{generalW}), it also peaks with the same value at $x=1$ where all buses bunch into one single unit.] However, the gradient here has an opposite sign since it is a larger $\Delta\theta$ from behind that improves the separation of the buses compared to a smaller $\Delta\theta$ from ahead. Fig.\ \ref{fig7}(a) shows the average waiting time for $N=2,3,4,5,6,7,8$ buses, respectively, serving $M=1$ bus stop in a loop, with Fig.\ \ref{fig7}(b) showing the corresponding situation with $M=12$ bus stops. Just like the previous section, additional bus stops bump up the waiting time since the bus has to spend some stoppage there for alighting/boarding. In these graphs, the simulation points are plotted with respect to implemented angles, which are generally smaller than the effective angles (analogous to looking-ahead in the previous section, where the implemented angle is larger than the effective angle). This leads to the shift between the simulation points and the analytical curves.

In comparison with looking immediately ahead, here looking immediately behind never allows any bunching. In fact, simulations show that the system does achieve a best average waiting time which is close to the theoretical minimum. In particular, the $N=8$ system has the best result, with sub-$0.1$ units of average waiting time. However, the range $\theta_0$ for implementing the no-boarding policy gets narrower with increasing $N$, in contrast to the implementation with respect to the bus immediately ahead where the range of $\theta_0$ grows with increasing $N$.

\section{No-boarding policy on buses with different natural frequencies}

In the previous two sections, we assumed that buses move with the same natural frequency. This allowed us to analytically calculate the average waiting time of the passengers waiting at the bus stop for a bus to arrive, complemented by simulations. Let us now test the no-boarding policy on a bus system where the buses move at different natural frequencies, as this is typical in human-driven buses.

As mentioned in Section 3B, the NTU campus bus system comprises $N$ buses serving $M=12$ reasonably staggered bus stops in a loop. The average periods of the buses around the loop without stopping are within a range of 12 to 18 minutes, or $f_i\in[0.93,1.39]$ mHz. Ref. \cite{Vee2019} found that for the system served by $N=2$ buses with $f_1=1.39$ mHz and $f_2=0.93$ mHz, the critical $k=k_c:=0.028$ marks a phase transition: 1) If $k<k_c$, the two buses are in the lull phase such that they periodically bunch. The bunching occurs periodically because the fast one is able to pull away, opening up the phase difference from the slower one, and subsequently lapping it. 2) On the other hand, if $k>k_c$, these two buses are in the busy phase where they are phase locked. Here, the phase locked pair is bunched permanently. Although the fast bus tries to pull away, the coupling strength at the bus stop is too large such that they bunch again --- repeatedly at every bus stop. For $N>2$, this value of $\bar{k}\sim k_c=0.028$ marks the approximate critical coupling strength where the system transitions from the lull phase with no permanent bunching occurring if $k<\bar{k}$, to the busy phase where at least one sustained phase-locked pair of buses exists if $k>\bar{k}$.

\begin{figure}
\centering
\includegraphics[width=16cm]{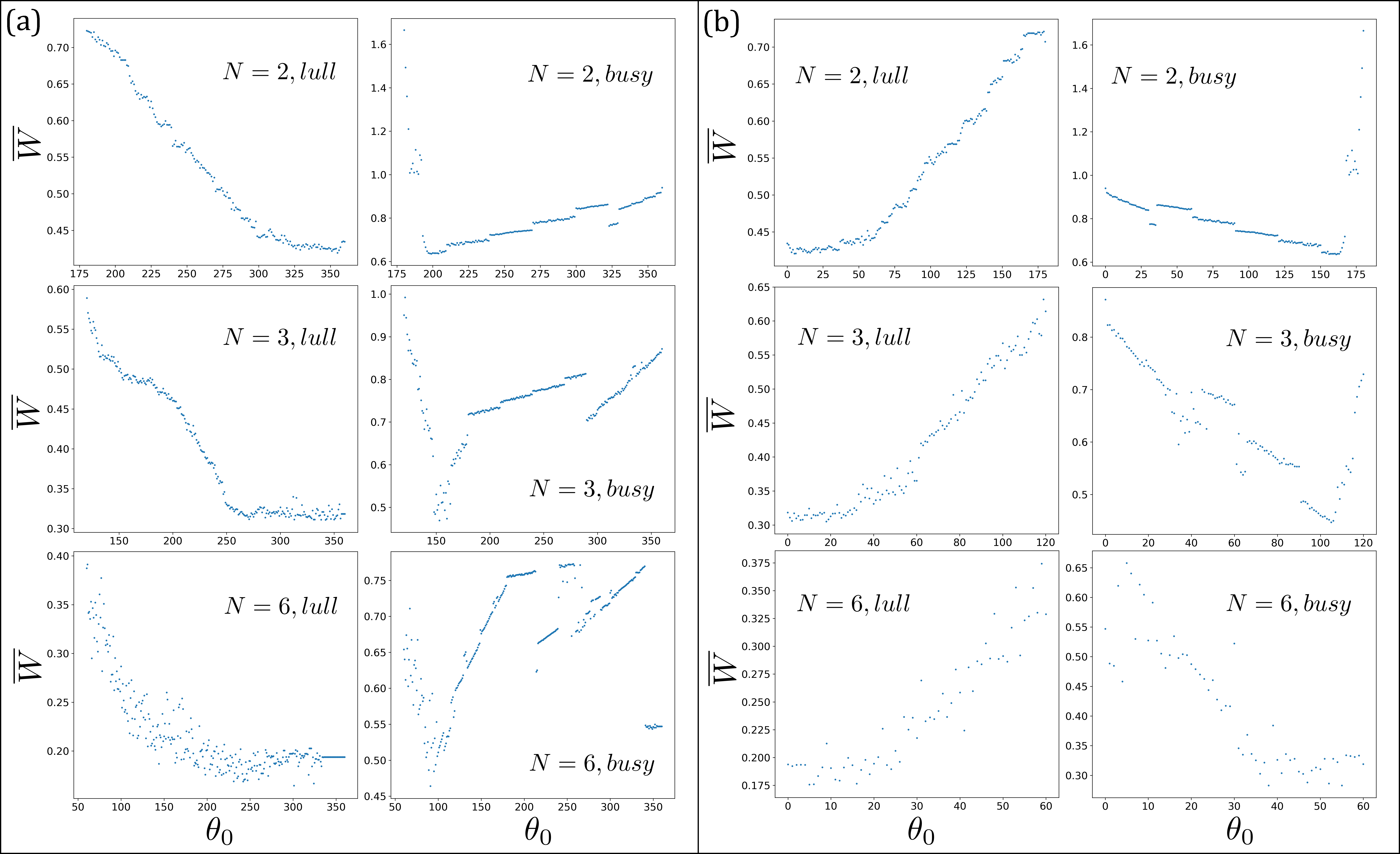}
\caption{(a) Simulation results for $N=2, 3, 6$ buses, respectively, serving $M=12$ bus stops in a loop. The no-boarding policy is applied by looking at the bus immediately ahead. Note that $\theta_0=360^\circ$ corresponds to no-implementation. The left column is in the lull phase where the no-boarding policy \emph{backfires} and increases the passenger waiting time, as compared to no implentation (or $360^\circ$). On the other hand, the right column is in the busy phase where the no-boarding policy correctly reduces the average waiting time as expected, by preventing the buses from forming permanent bunched clusters. (b) The corresponding simulation results as in (a), but with the no-boarding policy by looking at the bus immediately behind. Note that $\theta_0=0^\circ$ corresponds to no-implementation.}
\label{fig8}
\end{figure}

Let us now investigate the no-boarding policy applied to these situations in the NTU shuttle buses. We shall run simulations with the following setups:
\begin{enumerate}
\item A system with $N=2$ buses having frequencies $1.39$ and $0.93$ mHz, respectively, in the lull phase with $k=0.010$.
\item The system in (1), but in the busy phase with $k=0.030$.
\item A system with $N=3$ buses having frequencies $1.39$, $1.16$ and $0.93$ mHz, respectively, in the lull phase with $k=0.010$.
\item The system in (3), but in the busy phase with $k=0.040$.
\item A system with $N=6$ buses having frequencies $1.39$, $1.31$, $1.24$, $1.08$, $1.00$ and $0.93$ mHz, respectively, in the lull phase with $k=0.010$.
\item The system in (5), but in the busy phase with $k=0.063$.
\end{enumerate}
For each setup, we apply the no-boarding policy by looking immediately ahead [Fig.\ \ref{fig8}(a)], and repeat with the no-boarding policy by looking immediately behind [Fig.\ \ref{fig8}(b)]. Figs. \ref{fig8}(a) and \ref{fig8}(b) show the average waiting time in each case.

The no-boarding policy is effective, as expected from the analysis in the previous two sections, in the busy phase where otherwise the buses would experience sustained phase locking. However, the no-boarding policy \emph{backfires} when the system is in the lull phase. For the latter, the buses with frequency detuning only bunch periodically, with the fast one pulling away from the slow one after they leave the bus stop. The implementation of the no-boarding policy does not prevent the buses from bunching in the lull phase, because the fast bus can gradually defy the policy's efforts to maintain their phase difference due to its positive relative velocity. In fact, the no-boarding policy turns out to periodically make one of the buses end up not picking up anybody over a significant part of their time when the policy is enforced. See the video found here: \url{https://www.youtube.com/watch?v=SBNqvTr1AjQ} (please enable the caption to see the live description in the video), which annotates what happens when the $N=2$ bus system implements the no-boarding policy whenever $\Delta\theta>200^\circ$ by measuring the phase difference from the bus immediately ahead, during the lull phase ($k=0.010$). That video is also given as supplementary information.

\subsection{Deviation from the analytical curve in the busy phase}

It is interesting to note that unlike the previous setup with buses having identical natural frequency, buses with frequency detuning can potentially unbunch. Consequently for the no-boarding policy by looking at the bus immediately ahead, here the buses do not get stuck in the $(N-1)$-$1$ bus system since buses can unbunch. This is manifested by the $N=3$ and $N=6$ systems in the \emph{busy phase} [middle and bottom plots on the right column of Fig. \ref{fig8}(a)], where the simulation points do not lie on the expected curve for an $(N-1)$-$1$ system but we see that they manage to unbunch themselves to more efficient configurations. Furthermore for the $N=6$ system, having six buses with different frequencies turn out to be better than forcefully implementing the no-boarding policy over a wide range of implemented angle $\theta_0$. On the other hand, if the no-boarding policy is based on looking at the bus immediately behind [Fig.\ \ref{fig8}(b)], then the no-boarding policy generally improves the waiting time in the busy phase (right column), since it never allows any bunching in the first place.

\section{A simple general adaptive algorithm for dynamically determining \texorpdfstring{$\theta_0$}{angle}}\label{realworld}

For this section, we focus solely on the look-ahead version. A similar application would work with the look-behind version as well. In real bus systems, various sources of stochasticity would imply that the precise value of $\theta_\textrm{min}$, i.e. the lower bound to the angle to implement no-boarding $\theta_0$ (which would in principle minimise the average waiting time) does not quite exist or would be fluctuating with time. Therefore, we describe a simple algorithm for picking $\theta_0$ which is adaptive towards real-world stochasticity and varying demand levels $k_j$ for each bus stop. Before we present this algorithm, we first give some real-world parameters measured from the NTU campus shuttle buses \cite{Vee2019}. With this, we can use these parameters to define our simulation environment to mimic a realistic system, where we then implement the adaptive algorithm to dynamically pick $\theta_0$ which minimises the average waiting time.

\subsection{Passenger arrival rates for the 12 bus stops in the NTU loop campus shuttle bus service}

Ref. \cite{Vee2019} presented data on the NTU loop campus shuttle bus service for the \emph{Blue} route, measured over the full working week from 16th to 20th of April, 2018 (live data are found here: \url{https://baseride.com/maps/public/ntu/}). More specifically, there are two phases: 1) \emph{lull}, from 4 pm to 5 pm served by 3 buses; and 2) \emph{busy}, from 9 am to 10 am served by 6 or 7 buses. In that paper, loop averages are obtained where the values for $k$ are averaged over all $12$ bus stops. These loop averages produced data points which have relatively little noise, as the inhomogeneity amongst the bus stops are averaged away. The lull period for that week was measured to have a loop average of $k=0.024\pm0.004$ whilst the busy period at peak demand was measured to have a loop average of $k=0.065\pm0.017$. This peak demand was due to a selection of ten time series with the largest demand to obtain a representative value for the highest demand, where each time series is a tracking of one bus during that time interval. As discussed in the previous section, a phase transition between lull and busy was theoretically predicted to occur at $\bar{k}=0.028$, given buses with natural frequencies $f_i\in[0.93,1.39]$ mHz. This value $\bar{k}$ denotes the critical value of $k$ where the buses do not experience sustained bunching if $k<\bar{k}$ but clusters of phased-locked buses emerge when $k>\bar{k}$.

This loop average value of $k$ was obtained by fitting the equation $\tau=k\Delta t$ to the data points $(\Delta t,\tau)$, where the components are averages of the time headways $\Delta t_{ij}$ and stoppages $\tau_{ij}$ over all 12 bus stops, respectively. Note that in the NTU buses, passengers alight and board simultaneously via different doors, and so the stoppage is primarily due to the number of people accumulated during $\Delta t_{ij}$ at the bus stop, who are then boarding during $\tau_{ij}$. The error in $k$ is the square root of the mean squared deviation of each data point's corresponding $k$ from the best fit line's $k$. The relatively small errors in the loop average values of $k$ are consistent with the fact that they have reasonable fit with coefficient of determination $r^2$ values of $0.59$ and $0.77$ for the lull and busy periods, respectively. As mentioned earlier, the loop average irons out the inhomogeneity amongst the various bus stops' $k_j$.

\begin{table}
\centering
(a) \emph{Lull}
\vskip 0.5 cm
\begin{tabular}{|c|c|c|c|c|c|c|c|c|c|c|c|c|}
\hline
Bus stop & H4 & IC & SPMS & WKW & CEE & LWN & H3/16 & H14/15 & CH & H10/11 & H8 & H2\\
\hline
$k_j$ & 0.001 & 0.023 & 0.015 & 0.005 & 0.016 & 0.040 & 0.018 & 0.035 & 0.024 & 0.030 & 0.007 & 0.010\\
$\Delta k_j$ & 0.030 & 0.022 & 0.031 & 0.026 & 0.018 & 0.031 & 0.021 & 0.023 & 0.068 & 0.032 & 0.036 & 0.030\\
$\tau_{ij}$-intercept (s) & 23.3 & 17.3 & 17.4 & 19.3 & 10.9 & 22.0 & 13.2 & 9.9 & 25.8 & 29.3 & 23.9 & 19.8\\
$r^2$ & 0.00 & 0.19 & 0.07 & 0.01 & 0.16 & 0.41 & 0.22 & 0.39 & 0.18 & 0.13 & 0.01 & 0.04\\
\hline
\end{tabular}
\vskip 0.5 cm
(b) \emph{Busy}
\vskip 0.5 cm
\centering
\begin{tabular}{|c|c|c|c|c|c|c|c|c|c|c|c|c|}
\hline
Bus stop & H4 & IC & SPMS & WKW & CEE & LWN & H3/16 & H14/15 & CH & H10/11 & H8 & H2\\
\hline
$k_j$ & -0.019 & 0.063 & 0.026 & 0.033 & 0.008 & 0.027 & 0.067 & 0.001 & 0.006 & 0.063 & 0.003 & 0.031\\
$\Delta k_j$ & 0.082 & 0.096 & 0.077 & 0.068 & 0.053 & 0.077 & 0.052 & 0.152 & 0.069 & 0.077 & 0.093 & 0.071\\
$\tau_{ij}$-intercept (s) & 34.5 & 22.4 & 21.2 & 13.8 & 16.1 & 17.6 & 6.8 & 24.3 & 25.3 & 20.7 & 23.6 & 15.6\\
$r^2$ & 0.02 & 0.25 & 0.08 & 0.16 & 0.01 & 0.12 & 0.25 & 0.00 & 0.00 & 0.16 & 0.00 & 0.07\\
\hline
\end{tabular}
\caption{(a) The values of $k_j$ and their errors, for each of the $12$ bus stops in the NTU loop campus shuttle bus service, during the lull period (4 pm to 5 pm). These results are obtained from data measured over the entire working week of 16th to 20th of April, 2018. (b) The corresponding values during the busy period (9 am to 10 am).}\label{table1}
\end{table}

If we do this linear fit for each individual bus stop however, the measured values of $\Delta t_{ij}$ and $\tau_{ij}$ display a much greater error and deviation from the best fit line. The results are summarised in Table \ref{table1}. We find that the individual bus stop's error in $k_j$ are generally much larger than their loop average values for both the lull and busy periods, respectively. In fact, the variance is more pronounced during the busy period: Some bus stops like H14/15, CH, H8 have noticably small $k_j$, with H4 conspicuously having a negative $k_j$. However, their $\Delta k_j$ is very large with their $r^2$ values being close to $0$. This may be due to surges of people (students) collectively heading out to the bus stops at preferred times, say for example, to meet the 9.30 am lectures. The number of people going to the bus stops is thus not uniform over time, but experiences times when many people appear at the bus stops, as well as times when fewer people are there. The great stochasticity on individual bus stops may also be due to people crossing the road to take the bus service in the opposite direction when that bus arrives, since this is a loop service and people travelling antipodally (or near antipodally) may take a bus in either direction. The purpose of this crude linear fit is to set up a simulation environment (next subsection) to illustrate the adaptive algorithm on dynamically determining the best $\theta_0$ in a stochastic and non-stationary environment (two subsections later).

The $\tau_{ij}$-intercepts for each bus stop are primarily around the order of $+10$ seconds to $+30$ seconds, which is due to the fact that buses have to wait for clear traffic before rejoining the road (in fact, there are zebra crossings right in front of some of the bus stops, so some people would be crossing the road right after they alighted --- impeding the bus from departing) and that the positional data are only updated once in approximately every $10$ seconds (which implies that any delay would be recorded in multiples of $+10$ seconds). This is consistent with the loop average's $\tau$-intercept values being about $+20$ seconds, as discussed in Ref.\ \cite{Vee2019}.

\subsection{A simulated environment to model the NTU system}

For our simulated environment to model the NTU system, we assume that the buses are programmable such that they all move with the same natural period of $T=720$ seconds (excluding stoppages). We generate the rates of people arriving at each of the 12 bus stops according the their $k_j$ values from Table\ \ref{table1} for simulations on the lull and busy periods, respectively. To create some variations in $k_j$ with time, at every fixed time interval, we sample $k_{j,\textrm{sample}}$ for each bus stop according to a normal distribution with mean $k_j$ and standard deviation $\Delta k_j$. Note however that this may allow for negative values of $k_{j,\textrm{sample}}$ to be sampled, which is invalid and unphysical. To overcome this, we truncate the normal distribution to take only values from $0$ to $2k_j$ in order to maintain the distribution to center around $k_j$. (For H4 during the busy period with negative $k_j$, we take $|k_j|$ as the mean.) These sampled $k_{j,\textrm{sample}}$ will be taken as the values for $s_j:=$ average rate of people arrival at the respective bus stops, with $l:=$ loading/unloading rate set at 1 person per second. Once $k_{j,\textrm{sample}}$ have been sampled, the number of people arriving at bus stop $j$ is then determined by a Poisson distribution with $\lambda_\text{Poisson}=k_{j,\textrm{sample}}$. As mentioned, the values for $k_{j,\textrm{sample}}$ are resampled every fixed time interval to reflect the varying demand level with time, and this fixed interval is a hyperparameter for this model, i.e. one is free to set any value as desired for the model. We let this interval be the period of each bus $T=720$ seconds, i.e. $k_{j,\textrm{sample}}$ for each bus stop are resampled from their respective truncated normal distributions every $720$ seconds or $12$ minutes, which seems realistic.

Whilst the 12 bus stops are not perfectly staggered around the loop in the NTU campus, they are quite reasonably spaced out \cite{Vee2019}. We place them equally spaced out in our simulated environment. Also, we ignore traffic conditions and let the buses move with constant speed when they are not at a bus stop. These should not be considered as unrealistic simplifications, because of the fact that the simulation environment repetitiously draws out $k_{j,\textrm{sample}}$ from the truncated normal distribution every $T=720$ seconds. They are then further subjected to Poisson distributions to determine how many people actually arrive at each bus stop, thus already generating pretty high stochasticity in the simulated environment.

If buses move with identical natural speeds, they may not be able to unbunch once they are bunched at a bus stop. To overcome this so that the buses may continue to explore other values of $\theta_0$ instead of remaining bunched, we dictate in this simulated environment that if buses bunch at a bus stop, then they may randomly decide to leave. This mechanism thus allows buses with identical natural speeds to unbunch and carry on as if the system gets reset with bunched buses getting repositioned.

\subsection{A simple general adaptive algorithm for dynamically determining \texorpdfstring{$\theta_0$}{angle}}

Our proposed simple adaptive algorithm for selecting the angle to implement no-boarding $\theta_0$ motivated by the classical multi-armed bandit in reinforcement learning \cite{Sutton}. In this setup, each bus is an agent, and they only experience one state repeatedly. Each time, they can select one out of a set of actions, measure the reward, and subsequently arrive at the same state to select one action, ad infinitum. The set of actions comprises discrete integer values of $\theta_0\in(360^\circ/N, 360^\circ)$, where $N$ buses are serving the loop. This range of allowed $\theta_0$ would ensure that $\theta_0\leq360^\circ/N$ never gets picked, which would otherwise always implement no-boarding, as we have found in the analytical theory. Once $\theta_0$ is picked, this is the value where no-boarding would be implemented if $\Delta\theta>\theta_0$ (until the next time the $Q$-value gets updated and a new ${\theta_0}_\textrm{new}$ gets selected, described below). At the start of the run, a value of $\theta_0$ is randomly chosen from its allowed integer values. Each value of $\theta_0$ has a $Q$-value associated with it, also randomly initialised to some sufficiently high value. These $Q$-values represent the average waiting time of people at the bus stop for a bus to arrive, associated with that $\theta_0$. Hence, we aim to \emph{minimise} the average waiting time here, instead of typically \emph{maximising} the reward in reinforcement learning \cite{Sutton}. One may choose to let each bus (agent) have its own set of $Q$-values, or all buses share one single $Q$-table. Here, we adopt the simple setup with one shared $Q$-table.

When a bus arrives at a bus stop, it allows passengers who wish to alight there to do so. Once this has completed, this bus measures the phase difference $\Delta\theta$ from the bus immediately ahead of it. If $\Delta\theta\leq\theta_0$, then it proceeds to allow boarding until there is nobody left to board and leave, otherwise $\Delta\theta>\theta_0$ triggers the implementation of no boarding and the bus leaves. For people who are boarded, their waiting times are recorded so that the average waiting time is subsequently calculated. For the case where nobody is boarded and the bus just leaves, the waiting times thus far for people denied boarding at the bus stop are recorded instead. This ensures that the bus receives feedback on performance, even in cases where no-boarding is always implemented and nobody gets boarded. In reality, the information on how long a person has waited at a bus stop for a bus to arrive may be collected by a mobile app, where a person registers the intention to board a bus after just arriving at a bus stop. For the NTU campus shuttle bus system, perhaps a nifty way to do so instead, would be to install WiFi routers at bus stops since essentially students, staff, and anybody who regularly present themselves in NTU would access the university WiFi service. By installing WiFi routers at bus stops, not only does the university provide internet service for people spending time at bus stops, but the routers also count how many people are present at a bus stop waiting for a bus to arrive, \emph{automatically and in real time}. Of course, this is just an approximate data collection mechanism, since some people may have multiple devices or there may be guests who do not log in to the NTU WiFi automatically. Nevertheless, such an approximation should suffice to obtain a representative average waiting time which requires no effort from the passengers since their devices connect to the WiFi routers automatically.

After some fixed time interval $U$, the average waiting time $\overline{W}$ of all passengers waiting at the bus stop for a bus to arrive during this time interval is calculated. This ``reward'' $\overline{W}$ is updated to the $Q$-value associated with $\theta_0$ by the rule \cite{Sutton}:
\begin{align}\label{Qupdate}
Q_{\theta_0}\leftarrow Q_{\theta_0}+\alpha(\overline{W}-Q_{\theta_0}).
\end{align}
Here, $\alpha\in(0,1]$ is the learning rate for the $Q$-values. After this update, a new action is selected by a \emph{bounded} $\varepsilon$-greedy criterion: With a probability of $1-\varepsilon$, the bus picks a new ${\theta_0}_\textrm{new}$ corresponding to that whose $Q$-value is \emph{minimum}, otherwise a new ${\theta_0}_\textrm{new}$ is picked randomly from $[\theta_0-lower,\theta_0+upper]$. We choose $lower=30^\circ$ and $upper=5^\circ$ which bounds the $\varepsilon$-greedy exploration with a bias towards lower values of $\theta_0$, in order to keep the exploration controllable and headed towards the prior knowledge of the optimal $\theta_0$ to be near and above $\sim360^\circ/N$, based on the analytical theory in Section 3. We set $\alpha=0.2$ and $\varepsilon=0.2$. These are not decayed but kept constant, to be able to continually adapt to the non-stationary environment with varying $k_j$. Furthermore, we set $U=2T$, so the $Q$-value is updated every $1440$ seconds or $24$ minutes. It is the average waiting time $\overline{W}$ for passengers during this interval $U$ which is sent into Eq.\ (\ref{Qupdate}).

\subsection{Simulation results on the adaptive algorithm for determining \texorpdfstring{$\theta_0$}{angle} on the NTU system}

\begin{figure}
\centering
\includegraphics[width=16cm]{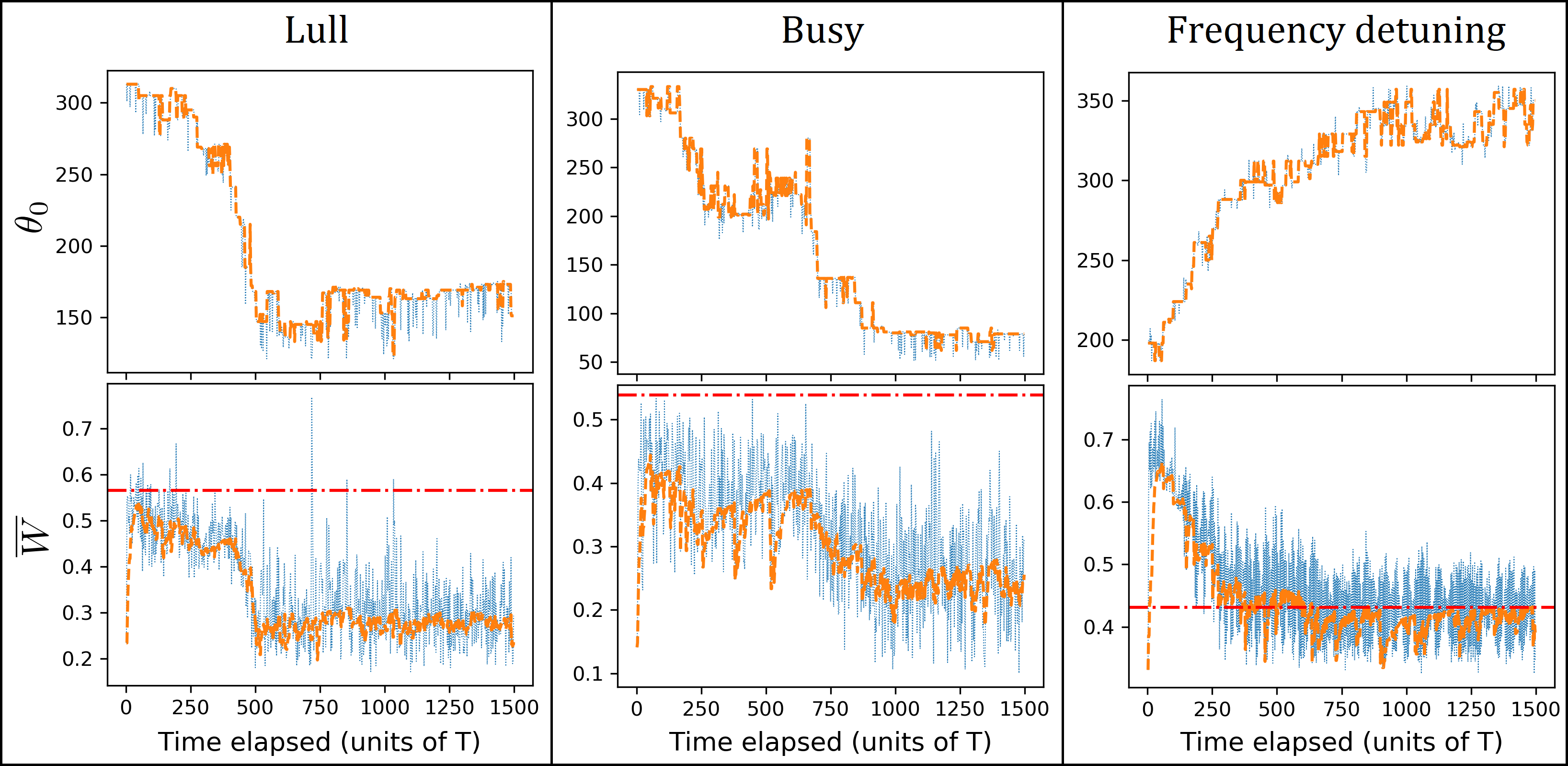}
\caption{Top row: Graphs of best and actual expected $\theta_0$ versus time for lull (left), busy (middle), and a lull case with frequency detuning (right). Bottom: Graphs of best and actual expected average waiting time versus time for lull (left), busy (middle), and a lull case with frequency detuning (right). In each plot, the actual curve is the thin dotted curve, whereas the best expected curve is the thick curve. The red horizontal line in the bottom graphs represents the average waiting time when the no-boarding policy is not implemented.}
\label{fig9}
\end{figure}

Fig.\ \ref{fig9} summarises the results of our adaptive algorithm to dynamically determine $\theta_0$, applied on the simulated environment of the NTU system, both in the lull (served by $N=3$ buses) and busy periods (served by $N=7$ buses), respectively. In each case, a large $\theta_0$ is initially chosen. As the simulation runs, various $\theta_0$ are explored according to the algorithm to minimise the average waiting time of people waiting at a bus stop for a bus to arrive. The plots (left column for lull, middle column for busy) show, as a function of time: 1) the best $\theta_0$ to implement, together with the actual $\theta_0$ taken (which with probability $\varepsilon$ is not the current best action, for exploration); 2) the best $Q$-value (representing the best expected average waiting time, based on the present $Q$-table), together with the actual average waiting time. Although we start with a large $\theta_0$, through reinforcement learning, the system is able to eventually seek the optimal $\theta_0$, and remain near that value since it has learnt that this has the best $Q$-value and occasionally bumps around to the next best $Q$-values when stochasticity drives up the average waiting time. This algorithm is also able to adapt to non-stationary situations, since $\varepsilon$ is kept fixed to maintain exploration.

Furthermore, when applied to a lull situation where buses have frequency detuning, the system is able to adapt and find the optimal $\theta_0$ to minimise waiting time. The right column in Fig. \ref{fig9} is the adaptive algorithm applied to ``setup (1)'' from the previous section, where there are $N=2$ buses with different natural frequencies, and each of the $M=12$ bus stops have $k=0.010$. The system is given an initially small $\theta_0\sim200^\circ$, but is able to explore and find the optimal $\theta_0\sim340^\circ$ which is slightly better than no implementation of the no-boarding policy [cf. top left plot in Fig.\ \ref{fig8}(a)]. Here, we use $lower=15^\circ,upper=15^\circ$ for the exploration, since we have no prior assumption about what the optimal $\theta_0$ should be.

The purpose of this description is to provide a means of picking an optimal $\theta_0$ under uncertain and stochastic real-world conditions, with some suggested hyperparameters like $\alpha=0.2,\varepsilon=0.2,lower=30^\circ,upper=5^\circ$ (or $lower=upper=15^\circ$ if no prior assumption is made), $ U=2T$. Whilst this prescription is intended to be general enough to be applicable to generic bus systems, when more specific details of a particular bus system are known, one may certainly fine-tune the hyperparameters and even make modifications (like giving each bus its own $Q$-table, instead of a commonly shared one) to improve this simple algorithm for specific bus systems.

\section{Discussion and concluding remarks}

Here is a summary of the results presented in this paper:
\begin{enumerate}
\item The construction of an analytical theory of how no-boarding significantly improves the bus system, as compared to the corresponding situation with no such implementation. This is for the case where buses have identical speed.
\item Extensive simulation results validate the theory.
\item The analytical theory comprises looking at the headway immediately \emph{ahead} or immediately \emph{behind}. The latter is generally better than the former.
\item Simulation results for the case where buses have different natural speeds show that no-boarding works only during the busy period, with an analytical derivation for this critical transition given in Ref.\ \cite{Chew2020}.
\item An adaptive algorithm is implemented on a simulation based on parameters measured from a real university loop shuttle bus service, illustrating how the no-boarding policy works in a real bus system.
\end{enumerate}
This paper has thus improved upon the previous studies on the no-boarding policy \cite{Del09,Del12,Zhao16} by providing a thorough theoretical framework that includes both alighting and boarding, elucidating the mechanism on how no-boarding improves and maintains the headways amongst buses, as well as exactly determining when no-boarding works (busy period with high demand) and does not (lull period with low demand).

Whilst the global average waiting time becomes more favorable, this comes with local cost as those denied boarding would experience increased waiting times. This social aspect definitely deserves further scrutiny. Sometimes passengers may have the urgent need for service, or certain weather conditions (heat, thunderstorm, heavy snow) are simply inconsiderately painful for these passengers to experience extended waiting time at the bus stop. It is arguably less of being a pain point when a passenger is on the bus, albeit slowly moving, compared to having to wait in the open at a bus stop \cite{Del09,Del12,Boardman18}. As a possible rectification, it may be a worthy consideration to allow passengers a choice on whether to obey the no-boarding implementation or defect and just continue boarding. A passenger defection against an implementation of the no-boarding policy would optimise their own local cost, at the expense of bumping up the global cost since this would raise the average waiting time for all passengers. These dynamics would seem to be rich and interesting, where the buses would have to co-evolve their no-boarding implementation strategies together with the passengers' strategies on whether to obey or defect (defecting potentially risks incurring a fine, for instance). These social and game theoretic aspects arising from the no-boarding policy would be studied systematically and reported elsewhere.

In particular, we have recently studied what happens when the no-boarding policy is not mandatorily enforced but defections are allowed \cite{Vee2019c}. This setup mimics the El Farol Bar problem \cite{Arthur94}, where each agent (player or commuter facing the no-boarding policy but allowed the option to defect) is given two choices and the winners are those in the ``minority group'', i.e. the smaller group is deemed as winners \cite{Challet97,Challet98,Savit99,Cavagna99,Manuca00,Moro04}. This is a typical problem on social-resource allocations like customers attending a bar with limited seats \cite{Arthur94}, a lunch/dinner crowd trying to get their meals from a number of possible restaurants \cite{Chakra07,Chakra09,Ghosh12,Chakra15}, the parking space problem \cite{Hanaki11,Kra19}, with even applications to financial markets where for instance if there are fewer buyers than sellers, then demand is weak and the price is low such that the buyers are ``winners'' (and vice versa) \cite{Zhang98,Challet00b,Challet01,Marsili01,Jeff01,Bou01,Challet01,Marsili02}. One of the key results of the minority game framework is that there exists a \emph{herding phase} whereby many agents may tend to choose the same action under certain conditions, which is bad in terms of resource allocation optimisation since many people not using implies excess resource whilst many people using implies overcrowding. For the no-boarding policy with allowance for defections, the herding phase with overwhelming defections would spell disaster and nullify the intended prevention of bus bunching. Nevertheless, we found \cite{Vee2019c} that since commuters face \emph{different} groups of other commuters each time a no-boarding policy is implemented, this differs from the classical minority game which assumes that the same group of agents play each other repeatedly and therefore, there is no herding behaviour. The conclusion here is that if a strict no-boarding policy is seen as too drastic, then an allowance for defections in a controlled manner (imposing a fine if there are too many defectors and giving rebates to cooperators to encourage cooperation) seems to be a viable way to improve the efficiency of the bus system by preventing bus bunching.

From a different perspective and in a separate paper, we have also carried out an interesting approach where we allow a simulation of buses to ``learn to be buses'' as well as discover strategies to improve the efficiency of the system --- \emph{without} human input or prior knowledge \cite{Vee2019d}. The idea there is similar to the sensational \emph{AlphaZero} programme which successfully dominated three different games of Go, Chess, and Shogi by \emph{reinforcement learning} \cite{AlphaZero}. In our work in Ref.\ \cite{Vee2019d} (not to be confused with Section \ref{realworld} in this paper), we allow each of the $N$ buses serving a loop of $M$ bus stops to either \emph{stay} or \emph{leave} whenever they are at a bus stop and there is nobody to alight (note that we do not allow for stop-skipping, as we do not wish to force commuters to stay on the bus for another round or having to get off at an earlier stop). The reward for the buses is that the system maintains a reasonably staggered configuration, analogous to the reward for \emph{AlphaZero} being to win the game. It turns out that, quite remarkably, reinforcement learning by interacting with other buses and without human input lead to the emergence of the \emph{no-boarding} and \emph{holding} strategies with a desirable effect of minimising the average waiting time of commuters at the bus stop for a bus to arrive. More specifically, the no-boarding strategy that they learn matches the analytical results derived in this paper --- including the upper bound strictly below $360^\circ/N$ where no-boarding is implemented [see Eq.\ (\ref{strictupperbound})]. Furthermore, the buses also learn cooperative strategies where they are able to unbunch (if they happen to bunch) and return to a more ideal configuration.

The no-boarding policy may be viewed as an entrainment mechanism of a system of self-oscillators, as pointed out and discussed in Ref. \cite{Vee2019}. Here, the entrainment mechanism for the bus system is triggered by $\Delta\theta$ exceeding a chosen $\theta_0$ (or $\Delta\theta<\theta_0$ in the look-behind version), and then a ``corrective force'' is applied by disallowing boarding on the slower bus. The effect is the \emph{staggered synchronisation} of buses where instead of the buses getting phase-locked with a phase difference of $0^\circ$, they are phased-locked with the ideal phase difference of $\sim360^\circ/N$ . This staggered synchronisation achieves significant reduction in the average waiting time of passengers. In fact, we have understood and shown mathematically how the no-boarding and holding strategies would \emph{create stable anti-bunched configurations} of buses serving a loop of bus stops, by comparing it to the local unidirectionally coupled Kuramoto oscillators which also possess stable staggered configurations if certain conditions are met \cite{Chew2020}. On top of that, this idea of synchronisation may also be useful in other systems, for example in the context of collaborative multicentre vehicle routing optimisation \cite{Wang17,Wang20}.

Although this study focuses on a loop where buses would continuously move without any start or end, the no-boarding policy is directly applicable to various topologies, viz. loops, linear bus routes with a start terminal and an end terminal, as well as bus routes with branches, etc., as the key idea is to disallow boarding when a slow bus is considered as being too slow. Thus, the main results of this work in terms of improvement of passengers' average waiting time is applicable to bus systems elsewhere apart from our NTU loop campus shuttle bus service, without the need for any sophisticated engineering or design of the bus routes and additional complex infrastructure.




\bibliography{Citation}

\begin{thebibliography}{70}%
\makeatletter
\providecommand \@ifxundefined [1]{%
 \@ifx{#1\undefined}
}%
\providecommand \@ifnum [1]{%
 \ifnum #1\expandafter \@firstoftwo
 \else \expandafter \@secondoftwo
 \fi
}%
\providecommand \@ifx [1]{%
 \ifx #1\expandafter \@firstoftwo
 \else \expandafter \@secondoftwo
 \fi
}%
\providecommand \natexlab [1]{#1}%
\providecommand \enquote  [1]{``#1''}%
\providecommand \bibnamefont  [1]{#1}%
\providecommand \bibfnamefont [1]{#1}%
\providecommand \citenamefont [1]{#1}%
\providecommand \href@noop [0]{\@secondoftwo}%
\providecommand \href [0]{\begingroup \@sanitize@url \@href}%
\providecommand \@href[1]{\@@startlink{#1}\@@href}%
\providecommand \@@href[1]{\endgroup#1\@@endlink}%
\providecommand \@sanitize@url [0]{\catcode `\\12\catcode `\$12\catcode
  `\&12\catcode `\#12\catcode `\^12\catcode `\_12\catcode `\%12\relax}%
\providecommand \@@startlink[1]{}%
\providecommand \@@endlink[0]{}%
\providecommand \url  [0]{\begingroup\@sanitize@url \@url }%
\providecommand \@url [1]{\endgroup\@href {#1}{\urlprefix }}%
\providecommand \urlprefix  [0]{URL }%
\providecommand \Eprint [0]{\href }%
\providecommand \doibase [0]{http://dx.doi.org/}%
\providecommand \selectlanguage [0]{\@gobble}%
\providecommand \bibinfo  [0]{\@secondoftwo}%
\providecommand \bibfield  [0]{\@secondoftwo}%
\providecommand \translation [1]{[#1]}%
\providecommand \BibitemOpen [0]{}%
\providecommand \bibitemStop [0]{}%
\providecommand \bibitemNoStop [0]{.\EOS\space}%
\providecommand \EOS [0]{\spacefactor3000\relax}%
\providecommand \BibitemShut  [1]{\csname bibitem#1\endcsname}%
\let\auto@bib@innerbib\@empty
\bibitem [{\citenamefont {Saw}\ \emph {et~al.}(2019{\natexlab{a}})\citenamefont
  {Saw}, \citenamefont {Chung}, \citenamefont {Quek}, \citenamefont {Pang},\
  and\ \citenamefont {Chew}}]{Vee2019}%
  \BibitemOpen
  \bibfield  {author} {\bibinfo {author} {\bibfnamefont {V.-L.}\ \bibnamefont
  {Saw}}, \bibinfo {author} {\bibfnamefont {N.~N.}\ \bibnamefont {Chung}},
  \bibinfo {author} {\bibfnamefont {W.~L.}\ \bibnamefont {Quek}}, \bibinfo
  {author} {\bibfnamefont {Y.~E.~I.}\ \bibnamefont {Pang}}, \ and\ \bibinfo
  {author} {\bibfnamefont {L.~Y.}\ \bibnamefont {Chew}},\ }\href {\doibase
  10.1038/s41598-019-43310-7} {\bibfield  {journal} {\bibinfo  {journal}
  {Scientific Reports}\ }\textbf {\bibinfo {volume} {9}},\ \bibinfo {pages}
  {6887} (\bibinfo {year} {2019}{\natexlab{a}})}\BibitemShut {NoStop}%
\bibitem [{\citenamefont {Newell}\ and\ \citenamefont
  {Potts}(1964)}]{Newell64}%
  \BibitemOpen
  \bibfield  {author} {\bibinfo {author} {\bibfnamefont {G.}~\bibnamefont
  {Newell}}\ and\ \bibinfo {author} {\bibfnamefont {R.}~\bibnamefont {Potts}},\
  }\href@noop {} {\bibfield  {journal} {\bibinfo  {journal} {2nd Australian
  Road Research Board}\ }\textbf {\bibinfo {volume} {2}},\ \bibinfo {pages}
  {388} (\bibinfo {year} {1964})}\BibitemShut {NoStop}%
\bibitem [{\citenamefont {Chapman}\ and\ \citenamefont
  {Michel}(1978)}]{Chapman78}%
  \BibitemOpen
  \bibfield  {author} {\bibinfo {author} {\bibfnamefont {R.~A.}\ \bibnamefont
  {Chapman}}\ and\ \bibinfo {author} {\bibfnamefont {J.~F.}\ \bibnamefont
  {Michel}},\ }\href {\doibase 10.1287/trsc.12.2.165} {\bibfield  {journal}
  {\bibinfo  {journal} {Transportation Science}\ }\textbf {\bibinfo {volume}
  {12}},\ \bibinfo {pages} {165} (\bibinfo {year} {1978})}\BibitemShut
  {NoStop}%
\bibitem [{\citenamefont {Powell}\ and\ \citenamefont
  {Sheffi}(1983)}]{Powell83}%
  \BibitemOpen
  \bibfield  {author} {\bibinfo {author} {\bibfnamefont {W.~B.}\ \bibnamefont
  {Powell}}\ and\ \bibinfo {author} {\bibfnamefont {Y.}~\bibnamefont
  {Sheffi}},\ }\href {\doibase 10.1287/trsc.17.4.376} {\bibfield  {journal}
  {\bibinfo  {journal} {Transportation Science}\ }\textbf {\bibinfo {volume}
  {17}},\ \bibinfo {pages} {376} (\bibinfo {year} {1983})}\BibitemShut
  {NoStop}%
\bibitem [{\citenamefont {Gershenson}\ and\ \citenamefont
  {Pineda}(2009)}]{Gers09}%
  \BibitemOpen
  \bibfield  {author} {\bibinfo {author} {\bibfnamefont {C.}~\bibnamefont
  {Gershenson}}\ and\ \bibinfo {author} {\bibfnamefont {L.~A.}\ \bibnamefont
  {Pineda}},\ }\href {\doibase 10.1371/journal.pone.0007292} {\bibfield
  {journal} {\bibinfo  {journal} {PLOS ONE}\ }\textbf {\bibinfo {volume} {4}},\
  \bibinfo {pages} {1} (\bibinfo {year} {2009})}\BibitemShut {NoStop}%
\bibitem [{\citenamefont {Bellei}\ and\ \citenamefont
  {Gkoumas}(2010)}]{Bell10}%
  \BibitemOpen
  \bibfield  {author} {\bibinfo {author} {\bibfnamefont {G.}~\bibnamefont
  {Bellei}}\ and\ \bibinfo {author} {\bibfnamefont {K.}~\bibnamefont
  {Gkoumas}},\ }\href {\doibase 10.1007/s12469-010-0024-7} {\bibfield
  {journal} {\bibinfo  {journal} {Public Transport}\ }\textbf {\bibinfo
  {volume} {2}},\ \bibinfo {pages} {269} (\bibinfo {year} {2010})}\BibitemShut
  {NoStop}%
\bibitem [{\citenamefont {Abkowitz}\ and\ \citenamefont
  {Engelstein}(1984)}]{Abk84}%
  \BibitemOpen
  \bibfield  {author} {\bibinfo {author} {\bibfnamefont {M.}~\bibnamefont
  {Abkowitz}}\ and\ \bibinfo {author} {\bibfnamefont {I.}~\bibnamefont
  {Engelstein}},\ }\href@noop {} {\bibfield  {journal} {\bibinfo  {journal}
  {Transportation Research Record: Journal of the Transportation Research
  Board}\ }\textbf {\bibinfo {volume} {961}},\ \bibinfo {pages} {1} (\bibinfo
  {year} {1984})}\BibitemShut {NoStop}%
\bibitem [{\citenamefont {Rossetti}\ and\ \citenamefont
  {Turitto}(1998)}]{Ros98}%
  \BibitemOpen
  \bibfield  {author} {\bibinfo {author} {\bibfnamefont {M.~D.}\ \bibnamefont
  {Rossetti}}\ and\ \bibinfo {author} {\bibfnamefont {T.}~\bibnamefont
  {Turitto}},\ }\href {\doibase https://doi.org/10.1016/S0965-8564(98)00019-6}
  {\bibfield  {journal} {\bibinfo  {journal} {Transportation Research Part A:
  Policy and Practice}\ }\textbf {\bibinfo {volume} {32}},\ \bibinfo {pages}
  {607 } (\bibinfo {year} {1998})}\BibitemShut {NoStop}%
\bibitem [{\citenamefont {Eberlein}\ \emph {et~al.}(2001)\citenamefont
  {Eberlein}, \citenamefont {Wilson},\ and\ \citenamefont
  {Bernstein}}]{Eber01}%
  \BibitemOpen
  \bibfield  {author} {\bibinfo {author} {\bibfnamefont {X.~J.}\ \bibnamefont
  {Eberlein}}, \bibinfo {author} {\bibfnamefont {N.~H.~M.}\ \bibnamefont
  {Wilson}}, \ and\ \bibinfo {author} {\bibfnamefont {D.}~\bibnamefont
  {Bernstein}},\ }\href {\doibase 10.1287/trsc.35.1.1.10143} {\bibfield
  {journal} {\bibinfo  {journal} {Transportation Science}\ }\textbf {\bibinfo
  {volume} {35}},\ \bibinfo {pages} {1} (\bibinfo {year} {2001})}\BibitemShut
  {NoStop}%
\bibitem [{\citenamefont {Hickman}(2001)}]{Hick01}%
  \BibitemOpen
  \bibfield  {author} {\bibinfo {author} {\bibfnamefont {M.~D.}\ \bibnamefont
  {Hickman}},\ }\href {http://www.jstor.org/stable/25768957} {\bibfield
  {journal} {\bibinfo  {journal} {Transportation Science}\ }\textbf {\bibinfo
  {volume} {35}},\ \bibinfo {pages} {215} (\bibinfo {year} {2001})}\BibitemShut
  {NoStop}%
\bibitem [{\citenamefont {Fu}\ and\ \citenamefont {Yang}(2002)}]{Fu02}%
  \BibitemOpen
  \bibfield  {author} {\bibinfo {author} {\bibfnamefont {L.}~\bibnamefont
  {Fu}}\ and\ \bibinfo {author} {\bibfnamefont {X.}~\bibnamefont {Yang}},\
  }\href {\doibase 10.3141/1791-02} {\bibfield  {journal} {\bibinfo  {journal}
  {Transportation Research Record}\ }\textbf {\bibinfo {volume} {1791}},\
  \bibinfo {pages} {6} (\bibinfo {year} {2002})}\BibitemShut {NoStop}%
\bibitem [{\citenamefont {Bin}\ \emph {et~al.}(2006)\citenamefont {Bin},
  \citenamefont {Zhongzhen},\ and\ \citenamefont {Baozhen}}]{Bin06}%
  \BibitemOpen
  \bibfield  {author} {\bibinfo {author} {\bibfnamefont {Y.}~\bibnamefont
  {Bin}}, \bibinfo {author} {\bibfnamefont {Y.}~\bibnamefont {Zhongzhen}}, \
  and\ \bibinfo {author} {\bibfnamefont {Y.}~\bibnamefont {Baozhen}},\ }\href
  {\doibase 10.1080/15472450600981009} {\bibfield  {journal} {\bibinfo
  {journal} {Journal of Intelligent Transportation Systems}\ }\textbf {\bibinfo
  {volume} {10}},\ \bibinfo {pages} {151} (\bibinfo {year} {2006})}\BibitemShut
  {NoStop}%
\bibitem [{\citenamefont {Daganzo}(2009)}]{Daganzo09}%
  \BibitemOpen
  \bibfield  {author} {\bibinfo {author} {\bibfnamefont {C.~F.}\ \bibnamefont
  {Daganzo}},\ }\href
  {http://www.sciencedirect.com/science/article/pii/S0191261509000484}
  {\bibfield  {journal} {\bibinfo  {journal} {Transportation Research Part B:
  Methodological}\ }\textbf {\bibinfo {volume} {43}},\ \bibinfo {pages} {913 }
  (\bibinfo {year} {2009})}\BibitemShut {NoStop}%
\bibitem [{\citenamefont {Cortés}\ \emph {et~al.}(2010)\citenamefont
  {Cortés}, \citenamefont {Sáez}, \citenamefont {Milla}, \citenamefont
  {Núñez},\ and\ \citenamefont {Riquelme}}]{Cor10}%
  \BibitemOpen
  \bibfield  {author} {\bibinfo {author} {\bibfnamefont {C.~E.}\ \bibnamefont
  {Cortés}}, \bibinfo {author} {\bibfnamefont {D.}~\bibnamefont {Sáez}},
  \bibinfo {author} {\bibfnamefont {F.}~\bibnamefont {Milla}}, \bibinfo
  {author} {\bibfnamefont {A.}~\bibnamefont {Núñez}}, \ and\ \bibinfo
  {author} {\bibfnamefont {M.}~\bibnamefont {Riquelme}},\ }\href {\doibase
  https://doi.org/10.1016/j.trc.2009.05.016} {\bibfield  {journal} {\bibinfo
  {journal} {Transportation Research Part C: Emerging Technologies}\ }\textbf
  {\bibinfo {volume} {18}},\ \bibinfo {pages} {757 } (\bibinfo {year}
  {2010})},\ \bibinfo {note} {applications of Advanced Technologies in
  Transportation: Selected papers from the 10th AATT Conference}\BibitemShut
  {NoStop}%
\bibitem [{\citenamefont {Cats}\ \emph {et~al.}(2011)\citenamefont {Cats},
  \citenamefont {Larijani}, \citenamefont {Koutsopoulos},\ and\ \citenamefont
  {Burghout}}]{Cats11}%
  \BibitemOpen
  \bibfield  {author} {\bibinfo {author} {\bibfnamefont {O.}~\bibnamefont
  {Cats}}, \bibinfo {author} {\bibfnamefont {A.~N.}\ \bibnamefont {Larijani}},
  \bibinfo {author} {\bibfnamefont {H.~N.}\ \bibnamefont {Koutsopoulos}}, \
  and\ \bibinfo {author} {\bibfnamefont {W.}~\bibnamefont {Burghout}},\ }\href
  {\doibase 10.3141/2216-06} {\bibfield  {journal} {\bibinfo  {journal}
  {Transportation Research Record}\ }\textbf {\bibinfo {volume} {2216}},\
  \bibinfo {pages} {51} (\bibinfo {year} {2011})}\BibitemShut {NoStop}%
\bibitem [{\citenamefont {Gershenson}(2011)}]{Gers11}%
  \BibitemOpen
  \bibfield  {author} {\bibinfo {author} {\bibfnamefont {C.}~\bibnamefont
  {Gershenson}},\ }\href {\doibase 10.1371/journal.pone.0021469} {\bibfield
  {journal} {\bibinfo  {journal} {PLOS ONE}\ }\textbf {\bibinfo {volume} {6}},\
  \bibinfo {pages} {1} (\bibinfo {year} {2011})}\BibitemShut {NoStop}%
\bibitem [{\citenamefont {Bartholdi}\ and\ \citenamefont
  {Eisenstein}(2012)}]{Bart12}%
  \BibitemOpen
  \bibfield  {author} {\bibinfo {author} {\bibfnamefont {J.~J.}\ \bibnamefont
  {Bartholdi}}\ and\ \bibinfo {author} {\bibfnamefont {D.~D.}\ \bibnamefont
  {Eisenstein}},\ }\href {\doibase 10.1016/j.trb.2011.11.001} {\bibfield
  {journal} {\bibinfo  {journal} {Transportation Research Part B:
  Methodological}\ }\textbf {\bibinfo {volume} {46}},\ \bibinfo {pages} {481 }
  (\bibinfo {year} {2012})}\BibitemShut {NoStop}%
\bibitem [{\citenamefont {Moreira-Matias}\ \emph {et~al.}(2016)\citenamefont
  {Moreira-Matias}, \citenamefont {Cats}, \citenamefont {Gama}, \citenamefont
  {Mendes-Moreira},\ and\ \citenamefont {de~Sousa}}]{Moreira16}%
  \BibitemOpen
  \bibfield  {author} {\bibinfo {author} {\bibfnamefont {L.}~\bibnamefont
  {Moreira-Matias}}, \bibinfo {author} {\bibfnamefont {O.}~\bibnamefont
  {Cats}}, \bibinfo {author} {\bibfnamefont {J.}~\bibnamefont {Gama}}, \bibinfo
  {author} {\bibfnamefont {J.}~\bibnamefont {Mendes-Moreira}}, \ and\ \bibinfo
  {author} {\bibfnamefont {J.~F.}\ \bibnamefont {de~Sousa}},\ }\href
  {http://www.sciencedirect.com/science/article/pii/S1568494616303118}
  {\bibfield  {journal} {\bibinfo  {journal} {Applied Soft Computing}\ }\textbf
  {\bibinfo {volume} {47}},\ \bibinfo {pages} {460 } (\bibinfo {year}
  {2016})}\BibitemShut {NoStop}%
\bibitem [{\citenamefont {Wang}\ \emph {et~al.}(2018)\citenamefont {Wang},
  \citenamefont {Chen}, \citenamefont {Chen}, \citenamefont {Cheng},\ and\
  \citenamefont {Lei}}]{Wang18}%
  \BibitemOpen
  \bibfield  {author} {\bibinfo {author} {\bibfnamefont {P.}~\bibnamefont
  {Wang}}, \bibinfo {author} {\bibfnamefont {X.}~\bibnamefont {Chen}}, \bibinfo
  {author} {\bibfnamefont {W.}~\bibnamefont {Chen}}, \bibinfo {author}
  {\bibfnamefont {L.}~\bibnamefont {Cheng}}, \ and\ \bibinfo {author}
  {\bibfnamefont {D.}~\bibnamefont {Lei}},\ }\href {\doibase
  10.1177/0361198118798722} {\bibfield  {journal} {\bibinfo  {journal}
  {Transportation Research Record}\ }\textbf {\bibinfo {volume} {0}},\ \bibinfo
  {pages} {0361198118798722} (\bibinfo {year} {2018})}\BibitemShut {NoStop}%
\bibitem [{\citenamefont {Li}\ \emph {et~al.}(1991)\citenamefont {Li},
  \citenamefont {Rousseau},\ and\ \citenamefont {Gendreau}}]{Li91}%
  \BibitemOpen
  \bibfield  {author} {\bibinfo {author} {\bibfnamefont {Y.}~\bibnamefont
  {Li}}, \bibinfo {author} {\bibfnamefont {J.-M.}\ \bibnamefont {Rousseau}}, \
  and\ \bibinfo {author} {\bibfnamefont {M.}~\bibnamefont {Gendreau}},\
  }\href@noop {} {\bibfield  {journal} {\bibinfo  {journal} {Proceedings of the
  thirty-third annual meeting, Transportation Research Forum}\ ,\ \bibinfo
  {pages} {157}} (\bibinfo {year} {1991})}\BibitemShut {NoStop}%
\bibitem [{\citenamefont {Eberlein}(1995)}]{Eber95}%
  \BibitemOpen
  \bibfield  {author} {\bibinfo {author} {\bibfnamefont {X.~J.}\ \bibnamefont
  {Eberlein}},\ }\href@noop {} {\bibfield  {journal} {\bibinfo  {journal} {PhD
  dissertation, Department of Civil and Environmental Engineering,
  Massachusetts Institute of Technology}\ } (\bibinfo {year}
  {1995})}\BibitemShut {NoStop}%
\bibitem [{\citenamefont {Fu}\ \emph {et~al.}(2003)\citenamefont {Fu},
  \citenamefont {Liu},\ and\ \citenamefont {Calamai}}]{Fu03}%
  \BibitemOpen
  \bibfield  {author} {\bibinfo {author} {\bibfnamefont {L.}~\bibnamefont
  {Fu}}, \bibinfo {author} {\bibfnamefont {Q.}~\bibnamefont {Liu}}, \ and\
  \bibinfo {author} {\bibfnamefont {P.}~\bibnamefont {Calamai}},\ }\href
  {\doibase 10.3141/1857-06} {\bibfield  {journal} {\bibinfo  {journal}
  {Transportation Research Record}\ }\textbf {\bibinfo {volume} {1857}},\
  \bibinfo {pages} {48} (\bibinfo {year} {2003})}\BibitemShut {NoStop}%
\bibitem [{\citenamefont {Sun}\ and\ \citenamefont {Hickman}(2005)}]{Sun05}%
  \BibitemOpen
  \bibfield  {author} {\bibinfo {author} {\bibfnamefont {A.}~\bibnamefont
  {Sun}}\ and\ \bibinfo {author} {\bibfnamefont {M.}~\bibnamefont {Hickman}},\
  }\href {\doibase 10.1080/15472450590934642} {\bibfield  {journal} {\bibinfo
  {journal} {Journal of Intelligent Transportation Systems}\ }\textbf {\bibinfo
  {volume} {9}},\ \bibinfo {pages} {91} (\bibinfo {year} {2005})}\BibitemShut
  {NoStop}%
\bibitem [{\citenamefont {Liu}\ \emph {et~al.}(2013)\citenamefont {Liu},
  \citenamefont {Yan}, \citenamefont {Qu},\ and\ \citenamefont
  {Zhang}}]{Liu13}%
  \BibitemOpen
  \bibfield  {author} {\bibinfo {author} {\bibfnamefont {Z.}~\bibnamefont
  {Liu}}, \bibinfo {author} {\bibfnamefont {Y.}~\bibnamefont {Yan}}, \bibinfo
  {author} {\bibfnamefont {X.}~\bibnamefont {Qu}}, \ and\ \bibinfo {author}
  {\bibfnamefont {Y.}~\bibnamefont {Zhang}},\ }\href {\doibase
  https://doi.org/10.1016/j.trc.2013.06.004} {\bibfield  {journal} {\bibinfo
  {journal} {Transportation Research Part C: Emerging Technologies}\ }\textbf
  {\bibinfo {volume} {35}},\ \bibinfo {pages} {46 } (\bibinfo {year}
  {2013})}\BibitemShut {NoStop}%
\bibitem [{\citenamefont {Furth}(1985)}]{Furth85}%
  \BibitemOpen
  \bibfield  {author} {\bibinfo {author} {\bibfnamefont {P.~G.}\ \bibnamefont
  {Furth}},\ }\href {\doibase 10.1287/trsc.19.1.13} {\bibfield  {journal}
  {\bibinfo  {journal} {Transportation Science}\ }\textbf {\bibinfo {volume}
  {19}},\ \bibinfo {pages} {13} (\bibinfo {year} {1985})}\BibitemShut {NoStop}%
\bibitem [{\citenamefont {Furth}\ and\ \citenamefont {Day}(1985)}]{Furth85b}%
  \BibitemOpen
  \bibfield  {author} {\bibinfo {author} {\bibfnamefont {P.}~\bibnamefont
  {Furth}}\ and\ \bibinfo {author} {\bibfnamefont {F.}~\bibnamefont {Day}},\
  }\href@noop {} {\bibfield  {journal} {\bibinfo  {journal} {Transportation
  Research Record}\ }\textbf {\bibinfo {volume} {1011}},\ \bibinfo {pages} {23}
  (\bibinfo {year} {1985})}\BibitemShut {NoStop}%
\bibitem [{\citenamefont {Eberlein}\ \emph {et~al.}(1998)\citenamefont
  {Eberlein}, \citenamefont {Wilson}, \citenamefont {Barnhart},\ and\
  \citenamefont {Bernstein}}]{Eber98}%
  \BibitemOpen
  \bibfield  {author} {\bibinfo {author} {\bibfnamefont {X.~J.}\ \bibnamefont
  {Eberlein}}, \bibinfo {author} {\bibfnamefont {N.~H.}\ \bibnamefont
  {Wilson}}, \bibinfo {author} {\bibfnamefont {C.}~\bibnamefont {Barnhart}}, \
  and\ \bibinfo {author} {\bibfnamefont {D.}~\bibnamefont {Bernstein}},\ }\href
  {http://www.sciencedirect.com/science/article/pii/S0191261597000131}
  {\bibfield  {journal} {\bibinfo  {journal} {Transportation Research Part B:
  Methodological}\ }\textbf {\bibinfo {volume} {32}},\ \bibinfo {pages} {77 }
  (\bibinfo {year} {1998})}\BibitemShut {NoStop}%
\bibitem [{\citenamefont {Lin}\ \emph {et~al.}(1995)\citenamefont {Lin},
  \citenamefont {Liang}, \citenamefont {Schonfeld},\ and\ \citenamefont
  {Larson}}]{Lin95}%
  \BibitemOpen
  \bibfield  {author} {\bibinfo {author} {\bibfnamefont {G.}~\bibnamefont
  {Lin}}, \bibinfo {author} {\bibfnamefont {P.}~\bibnamefont {Liang}}, \bibinfo
  {author} {\bibfnamefont {P.}~\bibnamefont {Schonfeld}}, \ and\ \bibinfo
  {author} {\bibfnamefont {R.}~\bibnamefont {Larson}},\ }\href@noop {}
  {\bibfield  {journal} {\bibinfo  {journal} {Final report for project
  MD-26-7002, University of Maryland}\ } (\bibinfo {year} {1995})}\BibitemShut
  {NoStop}%
\bibitem [{\citenamefont {Jara-D\'{i}az}\ and\ \citenamefont
  {Tirachini}(2013)}]{WideDoor}%
  \BibitemOpen
  \bibfield  {author} {\bibinfo {author} {\bibfnamefont {S.}~\bibnamefont
  {Jara-D\'{i}az}}\ and\ \bibinfo {author} {\bibfnamefont {A.}~\bibnamefont
  {Tirachini}},\ }\href@noop {} {\bibfield  {journal} {\bibinfo  {journal}
  {Journal of Tranport Economics and Policy}\ }\textbf {\bibinfo {volume}
  {47}},\ \bibinfo {pages} {91} (\bibinfo {year} {2013})}\BibitemShut {NoStop}%
\bibitem [{\citenamefont {Stewart}\ and\ \citenamefont
  {El-Geneidy}(2014)}]{Steward14}%
  \BibitemOpen
  \bibfield  {author} {\bibinfo {author} {\bibfnamefont {C.}~\bibnamefont
  {Stewart}}\ and\ \bibinfo {author} {\bibfnamefont {A.}~\bibnamefont
  {El-Geneidy}},\ }\href {\doibase 10.3141/2418-05} {\bibfield  {journal}
  {\bibinfo  {journal} {Transportation Research Record}\ }\textbf {\bibinfo
  {volume} {2418}},\ \bibinfo {pages} {39} (\bibinfo {year}
  {2014})}\BibitemShut {NoStop}%
\bibitem [{\citenamefont {El-Geneidy}\ \emph {et~al.}(2017)\citenamefont
  {El-Geneidy}, \citenamefont {van Lierop}, \citenamefont {Grisé},
  \citenamefont {Boisjoly}, \citenamefont {Swallow}, \citenamefont {Fordham},\
  and\ \citenamefont {Herrmann}}]{Geneidy17}%
  \BibitemOpen
  \bibfield  {author} {\bibinfo {author} {\bibfnamefont {A.}~\bibnamefont
  {El-Geneidy}}, \bibinfo {author} {\bibfnamefont {D.}~\bibnamefont {van
  Lierop}}, \bibinfo {author} {\bibfnamefont {E.}~\bibnamefont {Grisé}},
  \bibinfo {author} {\bibfnamefont {G.}~\bibnamefont {Boisjoly}}, \bibinfo
  {author} {\bibfnamefont {D.}~\bibnamefont {Swallow}}, \bibinfo {author}
  {\bibfnamefont {L.}~\bibnamefont {Fordham}}, \ and\ \bibinfo {author}
  {\bibfnamefont {T.}~\bibnamefont {Herrmann}},\ }\href
  {http://www.sciencedirect.com/science/article/pii/S0965856416306139}
  {\bibfield  {journal} {\bibinfo  {journal} {Transportation Research Part A:
  Policy and Practice}\ }\textbf {\bibinfo {volume} {99}},\ \bibinfo {pages}
  {114 } (\bibinfo {year} {2017})}\BibitemShut {NoStop}%
\bibitem [{\citenamefont {Tirachini}(2014)}]{Tirachini14}%
  \BibitemOpen
  \bibfield  {author} {\bibinfo {author} {\bibfnamefont {A.}~\bibnamefont
  {Tirachini}},\ }\href
  {http://www.sciencedirect.com/science/article/pii/S0965856413001900}
  {\bibfield  {journal} {\bibinfo  {journal} {Transportation Research Part A:
  Policy and Practice}\ }\textbf {\bibinfo {volume} {59}},\ \bibinfo {pages}
  {37 } (\bibinfo {year} {2014})}\BibitemShut {NoStop}%
\bibitem [{\citenamefont {Zhao}\ and\ \citenamefont {Zeng}(2008)}]{Zhao08}%
  \BibitemOpen
  \bibfield  {author} {\bibinfo {author} {\bibfnamefont {F.}~\bibnamefont
  {Zhao}}\ and\ \bibinfo {author} {\bibfnamefont {X.}~\bibnamefont {Zeng}},\
  }\href {\doibase https://doi.org/10.1016/j.ejor.2007.02.005} {\bibfield
  {journal} {\bibinfo  {journal} {European Journal of Operational Research}\
  }\textbf {\bibinfo {volume} {186}},\ \bibinfo {pages} {841 } (\bibinfo {year}
  {2008})}\BibitemShut {NoStop}%
\bibitem [{\citenamefont {Ceder}(2011)}]{Ceder11}%
  \BibitemOpen
  \bibfield  {author} {\bibinfo {author} {\bibfnamefont {A.}~\bibnamefont
  {Ceder}},\ }\href {\doibase https://doi.org/10.1016/j.sbspro.2011.08.005}
  {\bibfield  {journal} {\bibinfo  {journal} {Procedia - Social and Behavioral
  Sciences}\ }\textbf {\bibinfo {volume} {20}},\ \bibinfo {pages} {19 }
  (\bibinfo {year} {2011})},\ \bibinfo {note} {the State of the Art in the
  European Quantitative Oriented Transportation and Logistics Research – 14th
  Euro Working Group on Transportation \& 26th Mini Euro Conference \& 1st
  European Scientific Conference on Air Transport}\BibitemShut {NoStop}%
\bibitem [{\citenamefont {Tang}\ \emph {et~al.}(2018)\citenamefont {Tang},
  \citenamefont {Yang},\ and\ \citenamefont {Qi}}]{Tang18}%
  \BibitemOpen
  \bibfield  {author} {\bibinfo {author} {\bibfnamefont {J.}~\bibnamefont
  {Tang}}, \bibinfo {author} {\bibfnamefont {Y.}~\bibnamefont {Yang}}, \ and\
  \bibinfo {author} {\bibfnamefont {Y.}~\bibnamefont {Qi}},\ }\href@noop {}
  {\bibfield  {journal} {\bibinfo  {journal} {Physica A: Statistical Mechanics
  and its Applications}\ }\textbf {\bibinfo {volume} {512}},\ \bibinfo {pages}
  {745 } (\bibinfo {year} {2018})}\BibitemShut {NoStop}%
\bibitem [{\citenamefont {{Wang}}\ \emph {et~al.}(2017)\citenamefont {{Wang}},
  \citenamefont {{Zhang}}, \citenamefont {{Hu}}, \citenamefont {{Yang}},\ and\
  \citenamefont {{Lee}}}]{Wang17b}%
  \BibitemOpen
  \bibfield  {author} {\bibinfo {author} {\bibfnamefont {Y.}~\bibnamefont
  {{Wang}}}, \bibinfo {author} {\bibfnamefont {D.}~\bibnamefont {{Zhang}}},
  \bibinfo {author} {\bibfnamefont {L.}~\bibnamefont {{Hu}}}, \bibinfo {author}
  {\bibfnamefont {Y.}~\bibnamefont {{Yang}}}, \ and\ \bibinfo {author}
  {\bibfnamefont {L.~H.}\ \bibnamefont {{Lee}}},\ }\href {\doibase
  10.1109/TITS.2016.2644725} {\bibfield  {journal} {\bibinfo  {journal} {IEEE
  Transactions on Intelligent Transportation Systems}\ }\textbf {\bibinfo
  {volume} {18}},\ \bibinfo {pages} {2443} (\bibinfo {year}
  {2017})}\BibitemShut {NoStop}%
\bibitem [{\citenamefont {Delgado}\ \emph {et~al.}(2009)\citenamefont
  {Delgado}, \citenamefont {Muñoz}, \citenamefont {Giesen},\ and\
  \citenamefont {Cipriano}}]{Del09}%
  \BibitemOpen
  \bibfield  {author} {\bibinfo {author} {\bibfnamefont {F.}~\bibnamefont
  {Delgado}}, \bibinfo {author} {\bibfnamefont {J.~C.}\ \bibnamefont {Muñoz}},
  \bibinfo {author} {\bibfnamefont {R.}~\bibnamefont {Giesen}}, \ and\ \bibinfo
  {author} {\bibfnamefont {A.}~\bibnamefont {Cipriano}},\ }\href {\doibase
  10.3141/2090-07} {\bibfield  {journal} {\bibinfo  {journal} {Transportation
  Research Record}\ }\textbf {\bibinfo {volume} {2090}},\ \bibinfo {pages} {59}
  (\bibinfo {year} {2009})},\ \Eprint
  {http://arxiv.org/abs/https://doi.org/10.3141/2090-07}
  {https://doi.org/10.3141/2090-07} \BibitemShut {NoStop}%
\bibitem [{\citenamefont {Delgado}\ \emph {et~al.}(2012)\citenamefont
  {Delgado}, \citenamefont {Munoz},\ and\ \citenamefont {Giesen}}]{Del12}%
  \BibitemOpen
  \bibfield  {author} {\bibinfo {author} {\bibfnamefont {F.}~\bibnamefont
  {Delgado}}, \bibinfo {author} {\bibfnamefont {J.~C.}\ \bibnamefont {Munoz}},
  \ and\ \bibinfo {author} {\bibfnamefont {R.}~\bibnamefont {Giesen}},\ }\href
  {\doibase https://doi.org/10.1016/j.trb.2012.04.005} {\bibfield  {journal}
  {\bibinfo  {journal} {Transportation Research Part B: Methodological}\
  }\textbf {\bibinfo {volume} {46}},\ \bibinfo {pages} {1202 } (\bibinfo {year}
  {2012})}\BibitemShut {NoStop}%
\bibitem [{\citenamefont {Zhao}\ \emph {et~al.}(2016)\citenamefont {Zhao},
  \citenamefont {Lu}, \citenamefont {Liang},\ and\ \citenamefont
  {Liu}}]{Zhao16}%
  \BibitemOpen
  \bibfield  {author} {\bibinfo {author} {\bibfnamefont {S.~Z.}\ \bibnamefont
  {Zhao}}, \bibinfo {author} {\bibfnamefont {C.~X.}\ \bibnamefont {Lu}},
  \bibinfo {author} {\bibfnamefont {S.~D.}\ \bibnamefont {Liang}}, \ and\
  \bibinfo {author} {\bibfnamefont {H.~S.}\ \bibnamefont {Liu}},\ }\href
  {https://doi.org/10.1155/2016/8950209} {\bibfield  {journal} {\bibinfo
  {journal} {Mathematical Problems in Engineering}\ }\textbf {\bibinfo {volume}
  {2016}} (\bibinfo {year} {2016})}\BibitemShut {NoStop}%
\bibitem [{\citenamefont {Pikovsky}\ \emph {et~al.}(2003)\citenamefont
  {Pikovsky}, \citenamefont {Rosenblum},\ and\ \citenamefont {Kurths}}]{Syn03}%
  \BibitemOpen
  \bibfield  {author} {\bibinfo {author} {\bibfnamefont {A.}~\bibnamefont
  {Pikovsky}}, \bibinfo {author} {\bibfnamefont {M.}~\bibnamefont {Rosenblum}},
  \ and\ \bibinfo {author} {\bibfnamefont {J.}~\bibnamefont {Kurths}},\ }\href
  {http://www.cambridge.org/sg/academic/subjects/physics/nonlinear-science-and-fluid-dynamics/synchronization-universal-concept-nonlinear-sciences?format=PB&isbn=9780521533522}
  {\emph {\bibinfo {title} {Synchronization: A Universal Concept in Nonlinear
  Sciences}}},\ Cambridge Nonlinear Science Series\ (\bibinfo  {publisher}
  {Cambridge University Press},\ \bibinfo {address} {Cambridge},\ \bibinfo
  {year} {2003})\BibitemShut {NoStop}%
\bibitem [{\citenamefont {Chew}\ \emph {et~al.}(2020)\citenamefont {Chew},
  \citenamefont {Saw},\ and\ \citenamefont {Pang}}]{Chew2020}%
  \BibitemOpen
  \bibfield  {author} {\bibinfo {author} {\bibfnamefont {L.~Y.}\ \bibnamefont
  {Chew}}, \bibinfo {author} {\bibfnamefont {V.-L.}\ \bibnamefont {Saw}}, \
  and\ \bibinfo {author} {\bibfnamefont {Y.~E.~I.}\ \bibnamefont {Pang}},\
  }\href {https://arxiv.org/abs/1912.06470} {\emph {\bibinfo {title} {Stability
  of anti-bunched buses and local unidirectional Kuramoto oscillators}}}\
  (\bibinfo  {publisher} {To appear in a Book Chapter from the 15th
  International Conference on Dynamical Systems – Theory and Application in
  \L\'{o}d\'{z}, Poland, arXiv: 1912.06470},\ \bibinfo {year}
  {2020})\BibitemShut {NoStop}%
\bibitem [{NTU(2019)}]{NTUautobuses}%
  \BibitemOpen
  \href
  {https://media.ntu.edu.sg/NewsReleases/Pages/newsdetail.aspx?news=42801270-1948-431e-9502-1c725d513c83}
  {\emph {\bibinfo {title} {​NTU Singapore and Volvo unveil world’s first
  full size, autonomous electric bus}}}\ (\bibinfo  {publisher} {Nanyang
  Technological University Media Releases,
  https://media.ntu.edu.sg/NewsReleases/Pages/newsdetail.aspx?news=42801270-1948-431e-9502-1c725d513c83},\
  \bibinfo {year} {2019})\BibitemShut {NoStop}%
\bibitem [{\citenamefont {Begum}(2018)}]{NTUnews}%
  \BibitemOpen
  \bibfield  {author} {\bibinfo {author} {\bibfnamefont {S.}~\bibnamefont
  {Begum}},\ }\href {https://issuu.com/nanyangchronicle/docs/24.09_combined}
  {\bibfield  {journal} {\bibinfo  {journal} {9th of April, The Nanyang
  Chronicle}\ }\textbf {\bibinfo {volume} {24 (issue 9)}},\ \bibinfo {pages}
  {2} (\bibinfo {year} {2018})},\ \bibinfo {note} {``\emph{Need for higher
  shuttle bus frequency in the evenings: Survey}''}\BibitemShut {NoStop}%
\bibitem [{\citenamefont {Saw}\ \emph {et~al.}(2019{\natexlab{b}})\citenamefont
  {Saw}, \citenamefont {Vismara},\ and\ \citenamefont {Chew}}]{Vee2019d}%
  \BibitemOpen
  \bibfield  {author} {\bibinfo {author} {\bibfnamefont {V.-L.}\ \bibnamefont
  {Saw}}, \bibinfo {author} {\bibfnamefont {L.}~\bibnamefont {Vismara}}, \ and\
  \bibinfo {author} {\bibfnamefont {L.~Y.}\ \bibnamefont {Chew}},\ }\href
  {https://arxiv.org/abs/1911.03107} {\bibfield  {journal} {\bibinfo  {journal}
  {``Intelligent buses in a loop service: Emergence of \emph{no-boarding} and
  \emph{holding} strategies'', arXiv:1911.03107}\ } (\bibinfo {year}
  {2019}{\natexlab{b}})}\BibitemShut {NoStop}%
\bibitem [{\citenamefont {Sutton}\ and\ \citenamefont {Barto}(2018)}]{Sutton}%
  \BibitemOpen
  \bibfield  {author} {\bibinfo {author} {\bibfnamefont {R.}~\bibnamefont
  {Sutton}}\ and\ \bibinfo {author} {\bibfnamefont {A.}~\bibnamefont {Barto}},\
  }\href@noop {} {\emph {\bibinfo {title} {Reinforcement Learning: An
  Introduction}}}\ (\bibinfo  {publisher} {A Bradford Book},\ \bibinfo
  {address} {Cambridge, Massachusetts},\ \bibinfo {year} {2018})\BibitemShut
  {NoStop}%
\bibitem [{\citenamefont {Boardman}\ \emph {et~al.}(2018)\citenamefont
  {Boardman}, \citenamefont {Greenberg}, \citenamefont {Vining},\ and\
  \citenamefont {Weimer}}]{Boardman18}%
  \BibitemOpen
  \bibfield  {author} {\bibinfo {author} {\bibfnamefont {A.~E.}\ \bibnamefont
  {Boardman}}, \bibinfo {author} {\bibfnamefont {D.~H.}\ \bibnamefont
  {Greenberg}}, \bibinfo {author} {\bibfnamefont {A.~R.}\ \bibnamefont
  {Vining}}, \ and\ \bibinfo {author} {\bibfnamefont {D.~L.}\ \bibnamefont
  {Weimer}},\ }\href {\doibase 10.1017/9781108235594} {\emph {\bibinfo {title}
  {Cost-Benefit Analysis: Concepts and Practice}}},\ \bibinfo {edition} {5th}\
  ed.\ (\bibinfo  {publisher} {Cambridge University Press},\ \bibinfo {address}
  {Cambridge},\ \bibinfo {year} {2018})\BibitemShut {NoStop}%
\bibitem [{\citenamefont {Saw}\ and\ \citenamefont {Chew}(2019)}]{Vee2019c}%
  \BibitemOpen
  \bibfield  {author} {\bibinfo {author} {\bibfnamefont {V.-L.}\ \bibnamefont
  {Saw}}\ and\ \bibinfo {author} {\bibfnamefont {L.~Y.}\ \bibnamefont {Chew}},\
  }\href {https://arxiv.org/abs/1906.12110} {\bibfield  {journal} {\bibinfo
  {journal} {``No-boarding buses: Agents allowed to cooperate or defect'',
  Journal of Physics: Complexity, \emph{accepted}, arXiv:1906.12110}\ }
  (\bibinfo {year} {2019})}\BibitemShut {NoStop}%
\bibitem [{\citenamefont {Arthur}(1994)}]{Arthur94}%
  \BibitemOpen
  \bibfield  {author} {\bibinfo {author} {\bibfnamefont {W.~B.}\ \bibnamefont
  {Arthur}},\ }\href {http://www.jstor.org/stable/2117868} {\bibfield
  {journal} {\bibinfo  {journal} {The American Economic Review}\ }\textbf
  {\bibinfo {volume} {84}},\ \bibinfo {pages} {406} (\bibinfo {year}
  {1994})}\BibitemShut {NoStop}%
\bibitem [{\citenamefont {Challet}\ and\ \citenamefont
  {Zhang}(1997)}]{Challet97}%
  \BibitemOpen
  \bibfield  {author} {\bibinfo {author} {\bibfnamefont {D.}~\bibnamefont
  {Challet}}\ and\ \bibinfo {author} {\bibfnamefont {Y.-C.}\ \bibnamefont
  {Zhang}},\ }\href
  {http://www.sciencedirect.com/science/article/pii/S0378437197004196}
  {\bibfield  {journal} {\bibinfo  {journal} {Physica A: Statistical Mechanics
  and its Applications}\ }\textbf {\bibinfo {volume} {246}},\ \bibinfo {pages}
  {407 } (\bibinfo {year} {1997})}\BibitemShut {NoStop}%
\bibitem [{\citenamefont {Challet}\ and\ \citenamefont
  {Zhang}(1998)}]{Challet98}%
  \BibitemOpen
  \bibfield  {author} {\bibinfo {author} {\bibfnamefont {D.}~\bibnamefont
  {Challet}}\ and\ \bibinfo {author} {\bibfnamefont {Y.-C.}\ \bibnamefont
  {Zhang}},\ }\href
  {http://www.sciencedirect.com/science/article/pii/S037843719800260X}
  {\bibfield  {journal} {\bibinfo  {journal} {Physica A: Statistical Mechanics
  and its Applications}\ }\textbf {\bibinfo {volume} {256}},\ \bibinfo {pages}
  {514 } (\bibinfo {year} {1998})}\BibitemShut {NoStop}%
\bibitem [{\citenamefont {Savit}\ \emph {et~al.}(1999)\citenamefont {Savit},
  \citenamefont {Manuca},\ and\ \citenamefont {Riolo}}]{Savit99}%
  \BibitemOpen
  \bibfield  {author} {\bibinfo {author} {\bibfnamefont {R.}~\bibnamefont
  {Savit}}, \bibinfo {author} {\bibfnamefont {R.}~\bibnamefont {Manuca}}, \
  and\ \bibinfo {author} {\bibfnamefont {R.}~\bibnamefont {Riolo}},\ }\href
  {\doibase 10.1103/PhysRevLett.82.2203} {\bibfield  {journal} {\bibinfo
  {journal} {Phys. Rev. Lett.}\ }\textbf {\bibinfo {volume} {82}},\ \bibinfo
  {pages} {2203} (\bibinfo {year} {1999})}\BibitemShut {NoStop}%
\bibitem [{\citenamefont {Cavagna}(1999)}]{Cavagna99}%
  \BibitemOpen
  \bibfield  {author} {\bibinfo {author} {\bibfnamefont {A.}~\bibnamefont
  {Cavagna}},\ }\href {\doibase 10.1103/PhysRevE.59.R3783} {\bibfield
  {journal} {\bibinfo  {journal} {Phys. Rev. E}\ }\textbf {\bibinfo {volume}
  {59}},\ \bibinfo {pages} {R3783} (\bibinfo {year} {1999})}\BibitemShut
  {NoStop}%
\bibitem [{\citenamefont {Manuca}\ \emph {et~al.}(2000)\citenamefont {Manuca},
  \citenamefont {Li}, \citenamefont {Riolo},\ and\ \citenamefont
  {Savit}}]{Manuca00}%
  \BibitemOpen
  \bibfield  {author} {\bibinfo {author} {\bibfnamefont {R.}~\bibnamefont
  {Manuca}}, \bibinfo {author} {\bibfnamefont {Y.}~\bibnamefont {Li}}, \bibinfo
  {author} {\bibfnamefont {R.}~\bibnamefont {Riolo}}, \ and\ \bibinfo {author}
  {\bibfnamefont {R.}~\bibnamefont {Savit}},\ }\href
  {http://www.sciencedirect.com/science/article/pii/S037843710000100X}
  {\bibfield  {journal} {\bibinfo  {journal} {Physica A: Statistical Mechanics
  and its Applications}\ }\textbf {\bibinfo {volume} {282}},\ \bibinfo {pages}
  {559 } (\bibinfo {year} {2000})}\BibitemShut {NoStop}%
\bibitem [{\citenamefont {Moro}(2004)}]{Moro04}%
  \BibitemOpen
  \bibfield  {author} {\bibinfo {author} {\bibfnamefont {E.}~\bibnamefont
  {Moro}},\ }\href@noop {} {\emph {\bibinfo {title} {The Minority Game: An
  introductory guide}}},\ edited by\ \bibinfo {editor} {\bibfnamefont
  {E.}~\bibnamefont {Korutcheva}}\ and\ \bibinfo {editor} {\bibfnamefont
  {R.}~\bibnamefont {Cuerno}}\ (\bibinfo  {publisher} {Nova Science Publishers,
  Inc},\ \bibinfo {address} {New York},\ \bibinfo {year} {2004})\BibitemShut
  {NoStop}%
\bibitem [{\citenamefont {Chakrabarti}(2007)}]{Chakra07}%
  \BibitemOpen
  \bibfield  {author} {\bibinfo {author} {\bibfnamefont {B.~K.}\ \bibnamefont
  {Chakrabarti}},\ }\enquote {\bibinfo {title} {Kolkata restaurant problem as a
  generalised el farol bar problem},}\ in\ \href {\doibase
  10.1007/978-88-470-0665-2_18} {\emph {\bibinfo {booktitle} {Econophysics of
  Markets and Business Networks: Proceedings of the Econophys-Kolkata III}}},\
  \bibinfo {editor} {edited by\ \bibinfo {editor} {\bibfnamefont
  {A.}~\bibnamefont {Chatterjee}}\ and\ \bibinfo {editor} {\bibfnamefont
  {B.~K.}\ \bibnamefont {Chakrabarti}}}\ (\bibinfo  {publisher} {Springer
  Milan},\ \bibinfo {address} {Milano},\ \bibinfo {year} {2007})\ pp.\ \bibinfo
  {pages} {239--246}\BibitemShut {NoStop}%
\bibitem [{\citenamefont {Chakrabarti}\ \emph {et~al.}(2009)\citenamefont
  {Chakrabarti}, \citenamefont {Chakrabarti}, \citenamefont {Chatterjee},\ and\
  \citenamefont {Mitra}}]{Chakra09}%
  \BibitemOpen
  \bibfield  {author} {\bibinfo {author} {\bibfnamefont {A.~S.}\ \bibnamefont
  {Chakrabarti}}, \bibinfo {author} {\bibfnamefont {B.~K.}\ \bibnamefont
  {Chakrabarti}}, \bibinfo {author} {\bibfnamefont {A.}~\bibnamefont
  {Chatterjee}}, \ and\ \bibinfo {author} {\bibfnamefont {M.}~\bibnamefont
  {Mitra}},\ }\href {\doibase https://doi.org/10.1016/j.physa.2009.02.039}
  {\bibfield  {journal} {\bibinfo  {journal} {Physica A: Statistical Mechanics
  and its Applications}\ }\textbf {\bibinfo {volume} {388}},\ \bibinfo {pages}
  {2420 } (\bibinfo {year} {2009})}\BibitemShut {NoStop}%
\bibitem [{\citenamefont {Ghosh}\ \emph {et~al.}(2012)\citenamefont {Ghosh},
  \citenamefont {De~Martino}, \citenamefont {Chatterjee}, \citenamefont
  {Marsili},\ and\ \citenamefont {Chakrabarti}}]{Ghosh12}%
  \BibitemOpen
  \bibfield  {author} {\bibinfo {author} {\bibfnamefont {A.}~\bibnamefont
  {Ghosh}}, \bibinfo {author} {\bibfnamefont {D.}~\bibnamefont {De~Martino}},
  \bibinfo {author} {\bibfnamefont {A.}~\bibnamefont {Chatterjee}}, \bibinfo
  {author} {\bibfnamefont {M.}~\bibnamefont {Marsili}}, \ and\ \bibinfo
  {author} {\bibfnamefont {B.~K.}\ \bibnamefont {Chakrabarti}},\ }\href
  {\doibase 10.1103/PhysRevE.85.021116} {\bibfield  {journal} {\bibinfo
  {journal} {Phys. Rev. E}\ }\textbf {\bibinfo {volume} {85}},\ \bibinfo
  {pages} {021116} (\bibinfo {year} {2012})}\BibitemShut {NoStop}%
\bibitem [{\citenamefont {Chakraborti}\ \emph {et~al.}(2015)\citenamefont
  {Chakraborti}, \citenamefont {Challet}, \citenamefont {Chatterjee},
  \citenamefont {Marsili}, \citenamefont {Zhang},\ and\ \citenamefont
  {Chakrabarti}}]{Chakra15}%
  \BibitemOpen
  \bibfield  {author} {\bibinfo {author} {\bibfnamefont {A.}~\bibnamefont
  {Chakraborti}}, \bibinfo {author} {\bibfnamefont {D.}~\bibnamefont
  {Challet}}, \bibinfo {author} {\bibfnamefont {A.}~\bibnamefont {Chatterjee}},
  \bibinfo {author} {\bibfnamefont {M.}~\bibnamefont {Marsili}}, \bibinfo
  {author} {\bibfnamefont {Y.-C.}\ \bibnamefont {Zhang}}, \ and\ \bibinfo
  {author} {\bibfnamefont {B.~K.}\ \bibnamefont {Chakrabarti}},\ }\href
  {http://www.sciencedirect.com/science/article/pii/S0370157314003822}
  {\bibfield  {journal} {\bibinfo  {journal} {Physics Reports}\ }\textbf
  {\bibinfo {volume} {552}},\ \bibinfo {pages} {1 } (\bibinfo {year}
  {2015})}\BibitemShut {NoStop}%
\bibitem [{\citenamefont {Hanaki}\ \emph {et~al.}(2011)\citenamefont {Hanaki},
  \citenamefont {Kirman},\ and\ \citenamefont {Marsili}}]{Hanaki11}%
  \BibitemOpen
  \bibfield  {author} {\bibinfo {author} {\bibfnamefont {N.}~\bibnamefont
  {Hanaki}}, \bibinfo {author} {\bibfnamefont {A.}~\bibnamefont {Kirman}}, \
  and\ \bibinfo {author} {\bibfnamefont {M.}~\bibnamefont {Marsili}},\ }\href
  {http://www.sciencedirect.com/science/article/pii/S0167268110002325}
  {\bibfield  {journal} {\bibinfo  {journal} {Journal of Economic Behavior \&
  Organization}\ }\textbf {\bibinfo {volume} {77}},\ \bibinfo {pages} {382 }
  (\bibinfo {year} {2011})}\BibitemShut {NoStop}%
\bibitem [{\citenamefont {Krapivsky}\ and\ \citenamefont
  {Redner}(2019)}]{Kra19}%
  \BibitemOpen
  \bibfield  {author} {\bibinfo {author} {\bibfnamefont {P.~L.}\ \bibnamefont
  {Krapivsky}}\ and\ \bibinfo {author} {\bibfnamefont {S.}~\bibnamefont
  {Redner}},\ }\href {\doibase 10.1088/1742-5468/ab3a2a} {\bibfield  {journal}
  {\bibinfo  {journal} {Journal of Statistical Mechanics: Theory and
  Experiment}\ }\textbf {\bibinfo {volume} {2019}},\ \bibinfo {pages} {093404}
  (\bibinfo {year} {2019})}\BibitemShut {NoStop}%
\bibitem [{\citenamefont {Zhang}(1998)}]{Zhang98}%
  \BibitemOpen
  \bibfield  {author} {\bibinfo {author} {\bibfnamefont {Y.-C.}\ \bibnamefont
  {Zhang}},\ }\href@noop {} {\bibfield  {journal} {\bibinfo  {journal}
  {Europhysics News}\ }\textbf {\bibinfo {volume} {29}},\ \bibinfo {pages} {51}
  (\bibinfo {year} {1998})}\BibitemShut {NoStop}%
\bibitem [{\citenamefont {Challet}\ \emph {et~al.}(2000)\citenamefont
  {Challet}, \citenamefont {Marsili},\ and\ \citenamefont
  {Zhang}}]{Challet00b}%
  \BibitemOpen
  \bibfield  {author} {\bibinfo {author} {\bibfnamefont {D.}~\bibnamefont
  {Challet}}, \bibinfo {author} {\bibfnamefont {M.}~\bibnamefont {Marsili}}, \
  and\ \bibinfo {author} {\bibfnamefont {Y.-C.}\ \bibnamefont {Zhang}},\ }\href
  {https://www.sciencedirect.com/science/article/pii/S037843719900446X}
  {\bibfield  {journal} {\bibinfo  {journal} {Physica A: Statistical Mechanics
  and its Applications}\ }\textbf {\bibinfo {volume} {276}},\ \bibinfo {pages}
  {284 } (\bibinfo {year} {2000})}\BibitemShut {NoStop}%
\bibitem [{\citenamefont {Challet}\ \emph {et~al.}(2001)\citenamefont
  {Challet}, \citenamefont {Marsili},\ and\ \citenamefont {Zhang}}]{Challet01}%
  \BibitemOpen
  \bibfield  {author} {\bibinfo {author} {\bibfnamefont {D.}~\bibnamefont
  {Challet}}, \bibinfo {author} {\bibfnamefont {M.}~\bibnamefont {Marsili}}, \
  and\ \bibinfo {author} {\bibfnamefont {Y.-C.}\ \bibnamefont {Zhang}},\ }\href
  {http://www.sciencedirect.com/science/article/pii/S0378437101001030}
  {\bibfield  {journal} {\bibinfo  {journal} {Physica A: Statistical Mechanics
  and its Applications}\ }\textbf {\bibinfo {volume} {294}},\ \bibinfo {pages}
  {514 } (\bibinfo {year} {2001})}\BibitemShut {NoStop}%
\bibitem [{\citenamefont {Marsili}(2001)}]{Marsili01}%
  \BibitemOpen
  \bibfield  {author} {\bibinfo {author} {\bibfnamefont {M.}~\bibnamefont
  {Marsili}},\ }\href
  {http://www.sciencedirect.com/science/article/pii/S0378437101002850}
  {\bibfield  {journal} {\bibinfo  {journal} {Physica A: Statistical Mechanics
  and its Applications}\ }\textbf {\bibinfo {volume} {299}},\ \bibinfo {pages}
  {93 } (\bibinfo {year} {2001})}\BibitemShut {NoStop}%
\bibitem [{\citenamefont {Jefferies}\ \emph {et~al.}(2001)\citenamefont
  {Jefferies}, \citenamefont {Hart}, \citenamefont {Hui},\ and\ \citenamefont
  {Johnson}}]{Jeff01}%
  \BibitemOpen
  \bibfield  {author} {\bibinfo {author} {\bibfnamefont {P.}~\bibnamefont
  {Jefferies}}, \bibinfo {author} {\bibfnamefont {M.}~\bibnamefont {Hart}},
  \bibinfo {author} {\bibfnamefont {P.}~\bibnamefont {Hui}}, \ and\ \bibinfo
  {author} {\bibfnamefont {N.}~\bibnamefont {Johnson}},\ }\href {\doibase
  10.1007/s100510170228} {\bibfield  {journal} {\bibinfo  {journal} {The
  European Physical Journal B - Condensed Matter and Complex Systems}\ }\textbf
  {\bibinfo {volume} {20}},\ \bibinfo {pages} {493} (\bibinfo {year}
  {2001})}\BibitemShut {NoStop}%
\bibitem [{\citenamefont {Bouchaud}\ \emph {et~al.}(2001)\citenamefont
  {Bouchaud}, \citenamefont {Giardina},\ and\ \citenamefont
  {M\'{e}zard}}]{Bou01}%
  \BibitemOpen
  \bibfield  {author} {\bibinfo {author} {\bibfnamefont {J.-P.}\ \bibnamefont
  {Bouchaud}}, \bibinfo {author} {\bibfnamefont {I.}~\bibnamefont {Giardina}},
  \ and\ \bibinfo {author} {\bibfnamefont {M.}~\bibnamefont {M\'{e}zard}},\
  }\href {\doibase 10.1088/1469-7688/1/2/302} {\bibfield  {journal} {\bibinfo
  {journal} {Quantitative Finance}\ }\textbf {\bibinfo {volume} {1}},\ \bibinfo
  {pages} {212} (\bibinfo {year} {2001})},\ \Eprint
  {http://arxiv.org/abs/https://doi.org/10.1088/1469-7688/1/2/302}
  {https://doi.org/10.1088/1469-7688/1/2/302} \BibitemShut {NoStop}%
\bibitem [{\citenamefont {Marsili}\ and\ \citenamefont
  {Piai}(2002)}]{Marsili02}%
  \BibitemOpen
  \bibfield  {author} {\bibinfo {author} {\bibfnamefont {M.}~\bibnamefont
  {Marsili}}\ and\ \bibinfo {author} {\bibfnamefont {M.}~\bibnamefont {Piai}},\
  }\href {http://www.sciencedirect.com/science/article/pii/S0378437102008002}
  {\bibfield  {journal} {\bibinfo  {journal} {Physica A: Statistical Mechanics
  and its Applications}\ }\textbf {\bibinfo {volume} {310}},\ \bibinfo {pages}
  {234 } (\bibinfo {year} {2002})}\BibitemShut {NoStop}%
\bibitem [{\citenamefont {Silver}\ \emph {et~al.}(2018)\citenamefont {Silver},
  \citenamefont {Hubert}, \citenamefont {Schrittwieser}, \citenamefont
  {Antonoglou}, \citenamefont {Lai}, \citenamefont {Guez}, \citenamefont
  {Lanctot}, \citenamefont {Sifre}, \citenamefont {Kumaran}, \citenamefont
  {Graepel}, \citenamefont {Lillicrap}, \citenamefont {Simonyan},\ and\
  \citenamefont {Hassabis}}]{AlphaZero}%
  \BibitemOpen
  \bibfield  {author} {\bibinfo {author} {\bibfnamefont {D.}~\bibnamefont
  {Silver}}, \bibinfo {author} {\bibfnamefont {T.}~\bibnamefont {Hubert}},
  \bibinfo {author} {\bibfnamefont {J.}~\bibnamefont {Schrittwieser}}, \bibinfo
  {author} {\bibfnamefont {I.}~\bibnamefont {Antonoglou}}, \bibinfo {author}
  {\bibfnamefont {M.}~\bibnamefont {Lai}}, \bibinfo {author} {\bibfnamefont
  {A.}~\bibnamefont {Guez}}, \bibinfo {author} {\bibfnamefont {M.}~\bibnamefont
  {Lanctot}}, \bibinfo {author} {\bibfnamefont {L.}~\bibnamefont {Sifre}},
  \bibinfo {author} {\bibfnamefont {D.}~\bibnamefont {Kumaran}}, \bibinfo
  {author} {\bibfnamefont {T.}~\bibnamefont {Graepel}}, \bibinfo {author}
  {\bibfnamefont {T.}~\bibnamefont {Lillicrap}}, \bibinfo {author}
  {\bibfnamefont {K.}~\bibnamefont {Simonyan}}, \ and\ \bibinfo {author}
  {\bibfnamefont {D.}~\bibnamefont {Hassabis}},\ }\href {\doibase
  10.1126/science.aar6404} {\bibfield  {journal} {\bibinfo  {journal}
  {Science}\ }\textbf {\bibinfo {volume} {362}},\ \bibinfo {pages} {1140}
  (\bibinfo {year} {2018})}\BibitemShut {NoStop}%
\bibitem [{\citenamefont {Wang}\ \emph {et~al.}(2017)\citenamefont {Wang},
  \citenamefont {Ma}, \citenamefont {Li}, \citenamefont {Liu}, \citenamefont
  {Xu},\ and\ \citenamefont {Wang}}]{Wang17}%
  \BibitemOpen
  \bibfield  {author} {\bibinfo {author} {\bibfnamefont {Y.}~\bibnamefont
  {Wang}}, \bibinfo {author} {\bibfnamefont {X.}~\bibnamefont {Ma}}, \bibinfo
  {author} {\bibfnamefont {Z.}~\bibnamefont {Li}}, \bibinfo {author}
  {\bibfnamefont {Y.}~\bibnamefont {Liu}}, \bibinfo {author} {\bibfnamefont
  {M.}~\bibnamefont {Xu}}, \ and\ \bibinfo {author} {\bibfnamefont
  {Y.}~\bibnamefont {Wang}},\ }\href@noop {} {\bibfield  {journal} {\bibinfo
  {journal} {Journal of Cleaner Production}\ }\textbf {\bibinfo {volume}
  {144}},\ \bibinfo {pages} {203 } (\bibinfo {year} {2017})}\BibitemShut
  {NoStop}%
\bibitem [{\citenamefont {Wang}\ \emph {et~al.}(2020)\citenamefont {Wang},
  \citenamefont {Yuan}, \citenamefont {Guan}, \citenamefont {Xu}, \citenamefont
  {Wang}, \citenamefont {Wang},\ and\ \citenamefont {Liu}}]{Wang20}%
  \BibitemOpen
  \bibfield  {author} {\bibinfo {author} {\bibfnamefont {Y.}~\bibnamefont
  {Wang}}, \bibinfo {author} {\bibfnamefont {Y.}~\bibnamefont {Yuan}}, \bibinfo
  {author} {\bibfnamefont {X.}~\bibnamefont {Guan}}, \bibinfo {author}
  {\bibfnamefont {M.}~\bibnamefont {Xu}}, \bibinfo {author} {\bibfnamefont
  {L.}~\bibnamefont {Wang}}, \bibinfo {author} {\bibfnamefont {H.}~\bibnamefont
  {Wang}}, \ and\ \bibinfo {author} {\bibfnamefont {Y.}~\bibnamefont {Liu}},\
  }\href@noop {} {\bibfield  {journal} {\bibinfo  {journal} {Journal of Cleaner
  Production}\ }\textbf {\bibinfo {volume} {258}},\ \bibinfo {pages} {120590}
  (\bibinfo {year} {2020})}\BibitemShut {NoStop}%
\end{thebibliography}%

\begin{acknowledgments}
This work was supported by MOE AcRF Tier 1 (Grant No. RG93/15), the Joint WASP/NTU Programme (Project No. M4082189) and the DSAIR@NTU Grant (Project No. M4082418).
\end{acknowledgments}

\section*{Author Contributions}
V.-L.S and L.Y.C. designed research; V.-L.S. and L.Y.C. performed research; V.-L.S. collected and analysed data; V.-L.S. and L.Y.C. performed analytical study; V.-L.S. performed numerical simulations; V.-L.S. and L.Y.C wrote the manuscript; V.-L.S. prepared all figures.

\section*{Additional Information}
\subsection*{Supplementary information}
A video is included as supplementary information. It can be found here: \url{https://www.youtube.com/watch?v=SBNqvTr1AjQ} (Please enable caption to view the live description.)
\subsection*{NTU campus shuttle buses}
Live data on the NTU campus shuttle buses can be found here: \url{https://baseride.com/maps/public/ntu/}
\subsection*{Competing Interests}
The authors declare no competing interest.

\end{document}